\documentclass[twocolumn]{aastex631}
\usepackage{amsmath}
\usepackage{amssymb}
\usepackage{mathtools}
\usepackage{physics}
\usepackage{bm}
\usepackage{algorithm}
\usepackage{algpseudocode}
\usepackage{graphicx}
\usepackage{epstopdf}
\usepackage{ulem}
\newcommand{\ltsim}{\protect\raisebox{-0.5ex}{$\:\stackrel{\textstyle <}{\sim}\:$}}

\newcommand{\rev}[1]{{#1}}

\shorttitle{Multigrid Gravity Solver for Athena++}
\shortauthors{Tomida and Stone}
\begin{document}
\title{The Athena++ Adaptive Mesh Refinement Framework: Multigrid Solvers for Self-Gravity}

\author[0000-0001-8105-8113]{Kengo Tomida}\affiliation{Astronomical Institute, Tohoku University, Sendai, Miyagi, 980-8578, Japan\email{tomida@astr.tohoku.ac.jp}}
\author[0000-0001-5603-1832]{James M. Stone}\affiliation{School of Natural Sciences, Institute for Advanced Study, Princeton, NJ 08544, USA}

\begin{abstract}
We describe the implementation of multigrid solvers in the Athena++ adaptive mesh refinement (AMR) framework and their application to the solution of the Poisson equation for self-gravity. The new solvers are built on top of the AMR hierarchy and TaskList framework of Athena++ for efficient parallelization. We adopt a conservative formulation for the Laplacian operator that avoids artificial accelerations at level boundaries. Periodic, fixed, and zero-gradient boundary conditions are implemented, as well as open boundary conditions based on a multipole expansion. Hybrid parallelization using both MPI and OpenMP is adopted, and we present results of tests demonstrating the accuracy and scaling of the methods. On a uniform grid we show multigrid significantly outperforms methods based on FFTs, and requires only a small fraction of the compute time required by the (highly optimized) magnetohydrodynamic solver in Athena++.  As a demonstration of the capabilities of the methods, we present the results of a test calculation of magnetized protostellar collapse on an adaptive mesh.
\end{abstract}


\section{Introduction}
Modeling the (magneto-)hydrodynamics of astrophysical systems often requires inclusion of self-gravity, which necessitates solution of the Poisson equation along with the MHD conservation laws. Since it is elliptic, numerical solution of the Poisson equation requires quite different methods from the explicit time integration techniques typically used for the (hyperbolic) equations of motion. In addition, when diffusive processes such as viscosity, heat conduction, Ohmic resistivity, and radiation transport in the optically-thick regime (or with the flux-limited diffusion approximation) are included in the dynamics, implicit time integration methods are often needed in order to avoid overly-restrictive constraints on the timestep. Like \rev{the} solution of the Poisson equation, implicit integrators require matrix inversions that involve global communication.  Therefore it is of great interest to develop efficient solvers for such non-local physical processes that are scalable to a large number of parallel processes characteristic of modern high-performance computing systems.

There are a variety of widely adopted algorithms for the solution of the Poisson equation. For uniformly-spaced Cartesian grids, a discretization in Fourier space combined with the Fast Fourier Transform (FFT) is commonly used as it is robust and efficient \citep[e.g.][]{hockney,nr}. However, it is not compatible with non-uniform mesh spacing and curvilinear coordinates, is most efficient \rev{only with} periodic boundary conditions, and is not suitable for use with adaptive mesh refinement (AMR).  When AMR is used, sparse matrix solvers like the Conjugate-Gradient method and its variants \citep[e.g.][]{hockney,Hackbusch94,saad} can be adopted, but they are not trivial to implement on complicated grid structures as the matrix is irregular, and they may require robust preconditioners for efficiency. Another popular algorithm is the Tree method \citep{bhtree,flashgrav}, which is often used in particle-based simulations. While the Tree method is suitable for isolated (open) boundary conditions, it is not trivial to combine with other boundary conditions.

In this paper, we describe the implementation of grid-based solvers for the Poisson equation of self-gravity in the Athena++ code \citep{athenapp} that are based on the multigrid method \citep{fedorenko,multigrid}. In particular, we focus on the so-called geometric multigrid methods, which use multiple grids with different spatial resolutions. This method applies a classical iterative solver such as the Gauss-Seidel method to damp the error at different scales coherently. By construction, multigrid is compatible with both AMR and  a wide variety of different boundary conditions. Moreover, as long as communication costs are negligible, its computational complexity scales as $O(N)$, where $N$ is the total number of degrees of freedom. This is better than FFT and Tree algorithms, which both scale as $O(N\log N)$. In this paper, we show that this difference is significant at large $N$; multigrid is significantly more efficient than for example, FFT methods, in this regime.  We note that for the Poisson equation of self-gravity, the Fast Multipole Method \citep[FMM,][]{fmm} is also an efficient algorithm with $O(N)$.

The rest of the paper is organized as follows. In Section~2, we discuss the basics of multigrid methods. We describe our implementation of them in Section 3, and present test results in Section 4. In Section 5 we discuss and summarize our results. \rev{While the multigrid algorithms are compatible with general coordinate systems, we focus on uniformly-spaced Cartesian coordinates in this paper.}

\section{Multigrid Solvers}
\rev{In this section, we present basics of multigrid Poisson solvers. First, we describe the basic equations and their discretization in Section \ref{sec:basiceq}. Next, we discuss classical iterative solvers as a foundation of the multigrid algorithms in Section \ref{sec:classic}. Then we explain elements of the multigrid algorithms including the restriction and prolongation operators (Section \ref{sec:operators}), smoothing operator and coarsest-level solver (Section \ref{sec:smoother}), cycling algorithms (Sections \ref{sec:mgi}--\ref{sec:fas}), boundary conditions (Section \ref{sec:boundary}), and convergence criteria (Section \ref{sec:criterion}).}

\subsection{Basic Equations and Discretization} \label{sec:basiceq}
We discuss solution of the Poisson equation for self-gravity
\begin{eqnarray}
\nabla^2 \phi= 4\pi G \rho, \label{eq:poisson}\\
\bm{g} = -\nabla \phi, \label{eq:acceleration}
\end{eqnarray}
where $\phi$ is the gravitational potential, $\bm{g}$ the gravitational acceleration, $G$ the gravitational constant, and $\rho$ the mass density respectively. In three-dimensional (3D) Cartesian coordinates, the Laplacian operator becomes $\nabla^2=\frac{\partial^2}{\partial x^2}+\frac{\partial^2}{\partial y^2}+\frac{\partial^2}{\partial z^2}$. As Athena++ uses a cell-centered discretization for the hydrodynamic quantities, we use the same discretization for $\phi$ and $\rho$. 

For simplicity, in what follows the same uniform cell spacing $h$ is adopted in all the three directions (the extension to both non-uniform or curvilinear coordinates is straightforward). In this case, the discrete form of equation~(\ref{eq:poisson}) can be written 
\begin{eqnarray}
\phi_{i+1,j,k}+\phi_{i-1,j,k}+\phi_{i,j+1,k}+\phi_{i,j-1,k}+\phi_{i,j,k+1}+\phi_{i,j,k-1}\nonumber\\-6\phi_{i,j,k}= 4\pi G \rho_{i,j,k} h^2,\hspace{2em}\label{eq:discretized2}
\end{eqnarray}
where $i,j,k$ are the cell indices in the $x,y,z$-directions. Since Athena++ uses octree block-based AMR design, computations within each \texttt{MeshBlock} (domain decomposition unit in Athena++, see Section~\ref{sec:athenapp}) can be processed in locally-uniform grids and complexities due to AMR appear only in communications at level boundaries and refinement/derefinement procedures, as discussed in \rev{Section~\ref{sec:amrimp}}.

It is useful to introduce some notation. We use the potential and source term vectors $\bm{\phi}_h$ and $\bm{f}_h\equiv 4\pi G\bm{\rho}$ in which the values of all the cells discretized with the resolution of $h$ are stored as one dimensional vector. We define the Laplacian operator discretized with $h$, $L_h$. Then eq.(\ref{eq:discretized2}) reads 
\begin{eqnarray}
L_h\bm{\phi}_h=\bm{f}_h.\label{eq:discretized3}
\end{eqnarray}
We define the discretized error vector and true error vector as 
\begin{eqnarray}
\bm{\delta}_h(\bm{\phi}_h) &\equiv& \bm{\phi}_h^*-\bm{\phi}_h,\label{eq:error}\\
\bm{\epsilon}_h(\bm{\phi}_h) &\equiv& \bm{\phi}^*-\bm{\phi}_h,\label{eq:trueerror}
\end{eqnarray}
and the defect vector as
\begin{eqnarray}
\bm{d}_h(\bm{\phi}_h) &\equiv& \bm{f}_h-L_h\bm{\phi}_h,\label{eq:defect}
\end{eqnarray}
where $\bm{\phi}_h^*$ denotes the fully-converged discretized solution and $\bm{\phi}^*$ is the true solution of the original Poisson equation with infinite resolution. The converged solution satisfies $\bm{d}_h(\bm{\phi}_h^*) = \bm{0}$ and $\bm{\delta}_h(\bm{\phi}_h^*)=\bm{0}$. However, it should be noted that the discretized solution minimizes the true error $\bm{\epsilon}_h$ only to the order of the truncation error, which is proportional to $h^2$ in the case of a second-order accurate scheme.

The error and defect vectors satisfy the defect equation:
\begin{eqnarray}
L_h\bm{\delta}_h=\bm{d}_h.\label{eq:defecteq}
\end{eqnarray}
The Laplacian operator $L_h$ is a sparse matrix and its inversion $L_h^{-1}$ is not easy to find. Instead, we calculate an approximation $\hat{L}_h^{-1}$ to obtain the correction $\hat{\bm{\delta}}_{h}$ and then converge on the solution iteratively. We denote the approximate solution after the $n$-th iteration as $\phi_h^n$, while $\phi_h^0$ represents the initial guess.

\subsection{Classical Iterative Solvers}\label{sec:classic}
To be thorough, in this section we review basic concepts of classical iterative solvers such as the Jacobi, Gauss-Seidel (GS) and Successive Over-Relaxation (SOR) methods. These methods are not very practical by themselves, but they are important as the foundation of multigrid methods. There are various derivations of these methods \citep[e.g.][]{Hackbusch94}, here we present one \rev{physically motivated} \citep{nr}.

The Poisson equation can be reformulated as a time-dependent diffusion-like problem using the pseudo-time $\tau$:
\begin{equation}
\frac{\partial \phi}{\partial \tau}=\nabla^2 \phi-4\pi G \rho. \label{eq:pseudo}
\end{equation}
This is a linear diffusion equation with an additional source term. Because solutions to the Poisson equation are unique, we can start from any ``initial condition" and relax to the steady state solution of equation~(\ref{eq:pseudo}) where $\frac{\partial \phi}{\partial \tau}$ becomes zero, and the result will then be a solution of the Poisson equation.

In order to solve this diffusion-like equation numerically, the simplest approach is to apply the standard Forward Time Centered Space (FTCS) discretization \citep[e.g.][]{roache,nr} and use the largest possible timestep limited by the Courant-Friedrichs-Levy (CFL) condition, $\Delta \tau = h^2/2D$, where $D$ is the number of the spatial dimensions. The discretized equation in 3D is
\begin{eqnarray}
\phi^{n+1}_{i,j,k}=\frac{1}{6}\left(\phi^n_{i-1,j,k}+\phi^n_{i+1,j,k}+\phi^n_{i,j-1,k}+\phi^n_{i,j+1,k}\right.\nonumber\\+\left.\phi^n_{i,j,k-1}+\phi^n_{i,j,k+1}\right)-4\pi G \rho_{i,j,k} \frac{h^2}{6}, \label{eq:jacobi}
\end{eqnarray}
where $n$ is the index of the \rev{pseudo-time} or the iteration step. By repeating this procedure until $\phi$ reaches a steady state, we can obtain the numerical solution. This is the Jacobi method.

The timescale for the Jacobi methods to reach steady state is the diffusion time, $T\sim L^2/D$, where $L$ is the size of the system. This estimate tells us that short wavelength structures will diffuse on the shortest time scales, but long wavelength structures take longer to diffuse away. 
The number of iterations required for the whole system to reach steady state can be estimated as $N_{\rm iter}\sim T/\Delta \tau$\rev{$\propto N_x^2$} where $N_x=L/h$ is the number of the cells in each dimension. As each iteration costs $O(N_x^3)$ in 3D, the total computation complexity is $O(N_x^5)$. Compared to the explicit (magneto)hydrodynamic part, which is $O(N_x^3)$, this is very expensive.

Since we are interested only in the steady state solution of this problem, we are not strictly bound by the CFL condition and we can accelerate convergence by changing the iteration process. In the traditional (lexicographic) GS method, we replace eq.(\ref{eq:jacobi}) with
\begin{eqnarray}
\phi^{n+1}_{i,j,k}=\frac{1}{6}\left(\phi^{n+1}_{i-1,j,k}+\phi^{n}_{i+1,j,k}+\phi^{n+1}_{i,j-1,k}+\phi^{n}_{i,j+1,k}\right.\nonumber\\
+\left.\phi^{n+1}_{i,j,k-1}+\phi^{n}_{i,j,k+1}\right)-4\pi G \rho_{i,j,k} \frac{h^2}{6}. \label{eq:gs}
\end{eqnarray}
Here, $\phi^{n+1}$ is used on the right hand side for already updated cells. There are advantages and disadvantages in this method. The number of iterations required for convergence is only half of the Jacobi method \citep{hockney,Hackbusch94,nr}, and it does not require temporary memory storage for $\phi^{n+1}$ as we can directly update the value on the same memory. However, as it is obvious from the equation, this introduces anisotropy in the iteration process, and the dependency (i.e., we cannot update $\phi_{i,j,k}$ until $\phi_{i-1,j,k}, \phi_{i,j-1,k}, \phi_{i,j,k-1}$ are all updated) prevents vectorization and parallelization of the code. 

A variant of this is the Red-Black Gauss-Seidel (RBGS) method, in which a checkerboard pattern is used in the iteration. Here we label the even-numbered cells (i.e., $i+j+k$ = even) with red and the odd-numbered cells with black. We first update the red cells using the neighboring black cells yet to be updated, then the black cells using the red cells that are already updated.
\begin{eqnarray}
\phi^{n+1}_{i,j,k}=\frac{1}{6}\left(\phi^{n}_{i-1,j,k}+\phi^{n}_{i+1,j,k}+\phi^{n}_{i,j-1,k}+\phi^{n}_{i,j+1,k}\right.\nonumber\\+\left.\phi^{n}_{i,j,k-1}+\phi^{n}_{i,j,k+1}\right)-4\pi G \rho_{i,j,k} \frac{h^2}{6}\hspace{2em} \mathrm{(red)}\hspace{1em}\label{eq:rbgs1}\\
\phi^{n+1}_{i,j,k}=\frac{1}{6}\left(\phi^{n+1}_{i-1,j,k}+\phi^{n+1}_{i+1,j,k}+\phi^{n+1}_{i,j-1,k}+\phi^{n+1}_{i,j+1,k}\right.\nonumber\\+\left.\phi^{n+1}_{i,j,k-1}+\phi^{n+1}_{i,j,k+1}\right)-4\pi G \rho_{i,j,k} \frac{h^2}{6}\hspace{1em} \mathrm{(black)}\hspace{1em}\label{eq:rbgs2}
\end{eqnarray}
This method reduces anisotropy although red and black cells are not fully equivalent. Also, this method is suitable for parallelization as all the red or black cells can be updated at the same time without dependencies. However, it should be noted that this method is still incompatible with vectorization as it requires non-unit stride access to the memory. Both GS variants still cost $O(N_x^5)$.

We can even accelerate convergence further by SOR. In this method, we extrapolate each iteration by the over-relaxation parameter $\omega$. It can be combined with any of the classical solvers above, but in the case of the RBGS, the iteration procedure is as follows:
\begin{eqnarray}
\phi^{n+1}_{i,j,k}=\phi^n_{i,j,k}+\omega\left[\frac{1}{6}\left(\phi^{n}_{i-1,j,k}+\phi^{n}_{i+1,j,k}+\phi^{n}_{i,j-1,k}+\phi^{n}_{i,j+1,k}\right.\right.\nonumber\\
\left.\left.+\phi^{n}_{i,j,k-1}+\phi^{n}_{i,j,k+1}-6\phi^{n}_{i,j,k}\right)-4\pi G \rho_{i,j,k}\frac{h^2}{6}\right],\hspace{1.5em} \mathrm{(red)}\hspace{2em}\label{eq:rbsor1}\\
\phi^{n+1}_{i,j,k}=\phi^n_{i,j,k}+\omega\left[\frac{1}{6}\left(\phi^{n+1}_{i-1,j,k}+\phi^{n+1}_{i+1,j,k}+\phi^{n+1}_{i,j-1,k}+\phi^{n+1}_{i,j+1,k}\right.\right.\nonumber\\
\left.\left.+\phi^{n+1}_{i,j,k-1}+\phi^{n+1}_{i,j,k+1}-6\phi^{n+1}_{i,j,k}\right)-4\pi G \rho_{i,j,k}\frac{h^2}{6}\right].\hspace{0.5em} \mathrm{(black)}\hspace{2em}\label{eq:rbsor2}
\end{eqnarray}
This method is convergent for $0 < \omega < 2$. The optimal choice of $\omega$ depends on the problem, but
\begin{eqnarray}
\omega=\frac{2}{1+\pi/N_x},\label{eq:soromega}
\end{eqnarray}
is known to give significant acceleration for the Poisson equation for self-gravity if the domain is cubic. This reduces the number of iterations to $O(N_x)$, and therefore the total cost is $O(N_x^4)$ \citep{hockney,Hackbusch94,nr}.

\rev{It is known that SOR may not converge monotonically depending on the over-relaxation parameter $\omega$ and initial guess. To overcome this difficulty and improve the convergence behavior, the Chebyshev acceleration can be used. In this method, we use $\omega=1$ in the first iteration, and gradually increase it to the optimal value following a certain sequence. For further details, we refer readers to literature such as \citet{hockney,nr}, as we do not use this method in our multigrid solvers.}

\subsection{Basics of Multigrid}
The concept of multigrid methods is simple. By applying a classical iterative solver on coarse grids, long-wavelength modes damp faster as they behave as short-wavelength modes at the reduced resolution. Thus, we can use larger pseudo-timesteps on coarser grids and accelerate convergence to the steady state solution of the diffusion-like equation. The multigrid methods combine multiple layers of coarser grids consistently and damp both short- and long-wavelength modes coherently. In this section, we largely follow the discussion in \citet{multigrid}.

\rev{There are two different strategies in the multigrid methods. One is the multigrid iteration (MGI) method, in which we apply the multigrid method on the finest level repeatedly to improve the solution starting from a given initial guess. The other is the full multigrid (FMG) method, which starts from the coarsest grid, and constructs the finer solutions by applying the multigrid method using the coarser solution as the initial guess. We describe the multigrid iteration method in Section~\ref{sec:mgi} and the full multigrid method in Section~\ref{sec:fmg}}.

In what follows, we adopt the conventional terminology used with multigrid methods. Applying a classical iterative solver on a certain level is called ``smoothing". ``Restriction" means producing coarser grid data by averaging finer grid data, and ``prolongation" means producing finer grid data by interpolating coarser grid data. These are basic elements of the multigrid methods, and various combinations are possible. Here, we discuss our implementation in Athena++.

\rev{Hereafter, we also use notations that $l$ is the level in the multigrid hierarchy ($l=0$ is the coarsest level where the grid cannot be restricted anymore), $h_l$ the resolution on the level $l$ which satisfies $h_{l-1}=2h_l$, $L_l$ the Laplacian operator on level $l$, respectively.}

\begin{figure*}[t]
\begin{center}
\includegraphics[width=\textwidth]{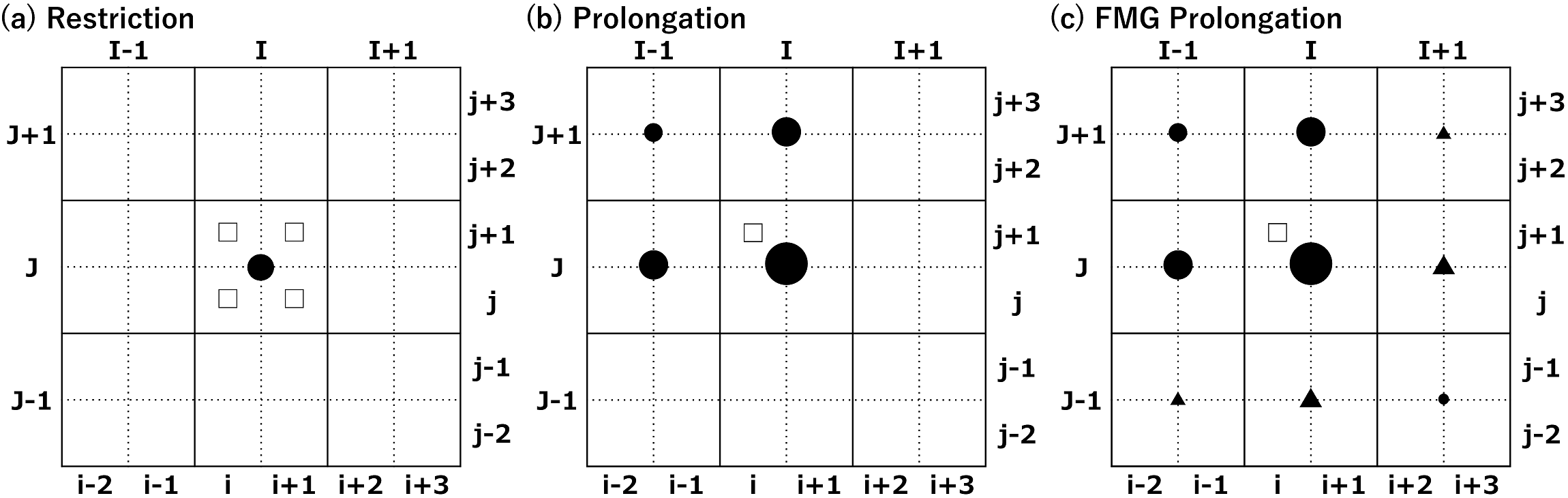}
\end{center}
\caption{2D representations of the interpolation stencils. The coarse grids are shown with solid lines while the fine grids are with dotted lines. The fine and coarse cell indices are marked with small and capital letters, respectively. (a) Restriction: filled \rev{circle} is calculated using the volume-weighted average of open \rev{squares}. (b) Prolongation: open square is calculated using trilinear (bilinear in 2D) interpolation using filled circles. (c) FMG prolongation:.open square is calculated using tricubic (bicubic in 2D) interpolation using filled circles and triangles. The sizes of the symbols \rev{qualitatively} correspond to the weighting coefficients while circles are positive and triangles are negative.}
\label{fig:stencil}
\end{figure*}

\subsubsection{Restriction and Prolongation Operators}\label{sec:operators}
\rev{First, we define the restriction and prolongation operators to transfer data between different levels.} As the potential and source term are discretized at cell-centers, we use the volume-weighted average for restriction and the trilinear interpolation for prolongation.

The restriction operator \rev{from level $l$ to $l-1$} is defined as 
\begin{eqnarray}
u_{l-1,I,J,K} = \left(I_l^{l-1} \bm{u}_l\right)_{I,J,K} \nonumber\\
= \frac{\sum_{a=0}^1\sum_{b=0}^1\sum_{c=0}^1 u_{l,i+a,j+b,k+c} \Delta V_{l,i+a,j+b,k+c}}{\sum_{a=0}^1\sum_{b=0}^1\sum_{c=0}^1\Delta V_{l,i+a,j+b,k+c}},\hspace{1ex}\label{eq:restriction}
\end{eqnarray}
where $u$ is either the potential $\phi$, \rev{the defect $d$,} or the source term $f$, $\Delta V_{l,i+a,j+b,k+c}$ the cell volume, and $I,J,K$ are the cell indices on the coarse level corresponding to the finer cells with $i,j,k$ and its adjacent cells.

The trilinear prolongation operator \rev{from level $l-1$ to $l$} is defined as
\begin{eqnarray}
u_{{l},i,j,k} = \left(I_{l-1}^{l} \bm{u}_{l-1}\right)_{i,j,k} \nonumber\\
=\sum_{a=-1}^{+1}\sum_{b=-1}^{+1}\sum_{c=-1}^{+1}w_{a,i}w_{b,j}w_{c,k}u_{h_{l-1},I+a,J+b,K+c},\label{eq:trilinear}
\end{eqnarray}
where \rev{$u$ is either the potential $\phi$ or the correction $\delta$}. $w_{n,m}$ takes $\frac{1}{4}, \frac{3}{4}, 0$ for $n=-1,0,1$ if $m$ is even (i.e., the finer cell is located on the ``lower" side in the coarser cell in each direction, see Figure~\ref{fig:stencil}), and $0, \frac{3}{4}, \frac{1}{4}$ for $n=-1,0,1$ if $m$ is odd (the finer cell is on the ``upper" side). These interpolation coefficients are derived assuming equally-spaced cubic cells in 3D Cartesian coordinates.

For the full multigrid method \rev{explained later in Section~\ref{sec:fmg}}, \rev{\citet{multigrid} recommended to use} a higher-order prolongation operator ${I'}_{l-1}^{l}$ to prolongate the potential, and we adopt the third-order tricubic interpolation for this purpose. It is obtained using eq.(\ref{eq:trilinear}) but with $w_{n,m}=\frac{5}{32}, \frac{30}{32}, \frac{-3}{32}$ for $n=-1,0,1$ and even $m$, and $w_{n,m}= \frac{-3}{32}, \frac{30}{32}, \frac{5}{32}$ for $n=-1,0,1$ and odd $m$.

The interpolation stencils and weighting coefficients of the restriction and prolongation operators in 2D are graphically shown in Figure~\ref{fig:stencil}.

\subsubsection{Smoothing Operator and Coarsest-Level Solver}\label{sec:smoother}
\rev{In the multigrid algorithms, we apply a smoothing operator $SMOOTH(\bm{u},L,\bm{f})$ to reduce the error on each level. This operator receives the potential or defect vector, the linear operator and the source term vector as the input arguments, and returns the smoothed potential or defect vector. In our implementation, we adopt the RBGS smoother with the over-relaxation factor of $\omega = 1.15$ as eqs.(\ref{eq:rbsor1}) and (\ref{eq:rbsor2}). This value of $\omega$ is chosen to damp high-frequency errors faster, and is different from the optimal value for the SOR solver in eq.(\ref{eq:soromega}) \citep{yavneh1996, multigrid}.

On the coarsest level, we apply the coarsest-level solver $COARSEST(\bm{u},L,\bm{f})$. The input arguments are the same as $SMOOTH()$, and it returns the solution on the coarsest level. If the grid can be reduced to $1\times1\times1$ cell on the coarsest level $l=0$, it is trivial to calculate the exact solution because the discretized Poisson equation (\ref{eq:discretized2}) is reduced to 
\begin{eqnarray}
u_{0,1,1,1} = \frac{1}{6}(u_{0,0,1,1}+u_{0,1,0,1}+u_{0,1,1,0}+u_{0,1,1,2}\nonumber\\+u_{0,1,2,1}+u_{0,2,1,1}-h_0^2f_{0,1,1,1}),
\end{eqnarray}
and all $u$ on the right hand side are given from boundary conditions. On the other hand, if the coarsest level is not reduced to a single cell but contains $N_{0,x}\times N_{0,y}\times N_{0,z}$ cells, we apply the RBGS smoother $N_{0,{\rm max}}\equiv{\rm max}(N_{0,x},N_{0,y},N_{0,z})$ times instead, which is sufficient to obtain a good solution on the coarsest level.}

\subsubsection{Multigrid Iteration}\label{sec:mgi}
\begin{figure*}[t]
\begin{center}
\includegraphics[width=\textwidth]{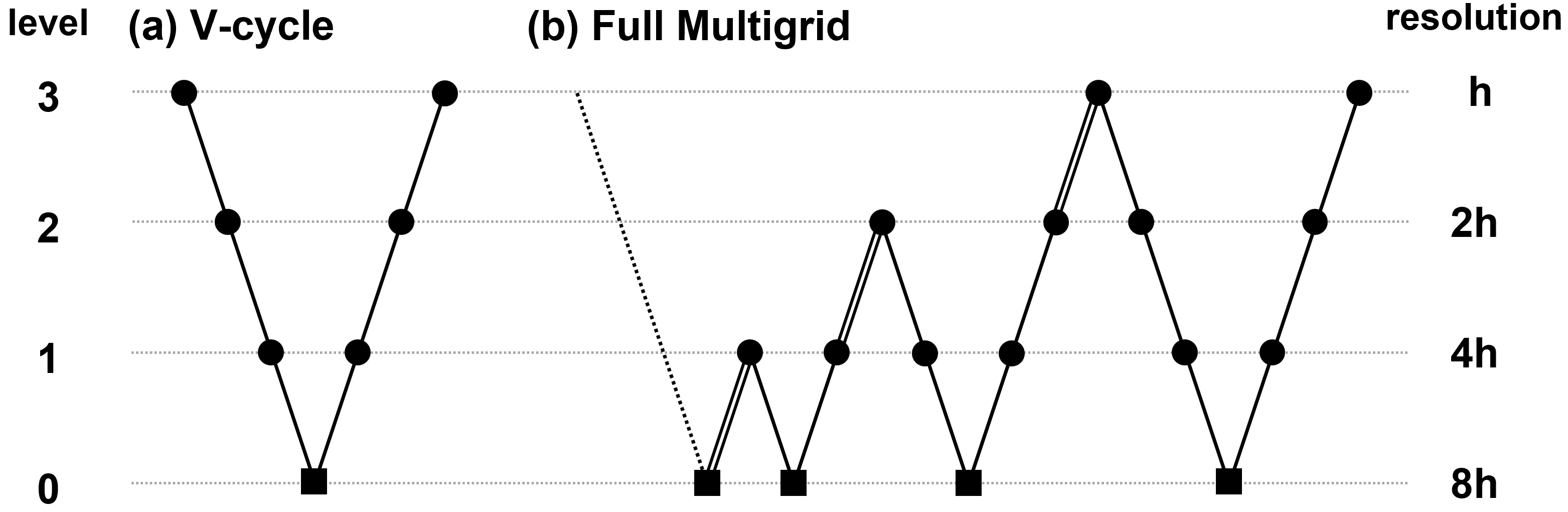}
\end{center}
\caption{Schematic graphs of the multigrid algorithms with four levels. (a) The multigrid iteration method with the V(1,1) cycles. (b) The full multigrid method with the V(1,1) cycles. \\
Legends: Filled circles ($\bullet$): smoothing operations. Filled squares ($\blacksquare$): obtaining the coarsest-grid solution. Downward solid segments: restriction. Upward solid segments: prolongation. Downward dotted line: restriction operation at the beginning of the full multigrid algorithm. Upward double-solid segments: FMG prolongation.}
\label{fig:cycles}
\end{figure*}
We first describe the multigrid iteration scheme. There are various cycling procedures \rev{such as V-, W-, and F-cycles, and here} we implement the simplest V-cycle in Athena++. Using pseudocode notation, one cycle of this scheme is described as follows.\\
{\bf Algorithm 1:} V-cycle multigrid iteration
\begin{algorithmic}
\Procedure{$MULTIGRID$}{$\bm{\phi}_{l}^n,L_{l},\bm{f}_{l},l$}
\State 1. Presmooth: $\bm{\bar{\phi}}_{l}^n=SMOOTH^M(\bm{\phi}_{l}^n,L_{l},\bm{f}_{l})$.
\State 2. Compute defect: $\bm{d}_{l}^n = \bm{f}_{l}-L_{l}\bm{\bar{\phi}}_{l}^n$.
\State 3. Restrict defect: $\bm{d}_{l-1}^n=I_{l}^{l-1}\bm{d}_{l}^n$.
\State 4. Calculate coarse-grid correction:
\If{$l = 1$}
  \State 4.1 Solve the coarsest level: \State \hspace{2ex}$\hat{\bm{\delta}}_{0}^n=COARSEST(\bm{0},L_0,\bm{d}_0^n)$.
\Else
  \State 4.2 Apply $MULTIGRID$ recursively: \State \hspace{2ex}$\hat{\bm{\delta}}_{l-1}^n=MULTIGRID(\bm{0},L_{l-1},\bm{d}_{l-1}^n,l-1)$.
\EndIf
\State 5. Prolongate correction: $\hat{\bm{\delta}}_{l}^n=I_{l-1}^{l}\hat{\bm{\delta}}_{l-1}^n$.
\State 6. Apply correction: ${\bm{\phi}'}_{l}^n=\bm{\bar{\phi}}_{l}^n+\hat{\bm{\delta}}_{l}^n$.
\State 7. Postsmooth: $\bm{\phi}_{l}^{n+1}=SMOOTH^N({\bm{\phi}'}_{l}^n,L_{l},\bm{f}_{l})$.
\State {\bf return} $\bm{\phi}_{l}^{n+1}$.
\EndProcedure
\end{algorithmic}
Here, $I_l^{l-1}$ and $I_{l-1}^{l}$ are the restriction and prolongation operators \rev{defined in Section~\ref{sec:operators}, $SMOOTH()$ and $COARSEST()$ the smoothing operator and coarsest-level solver defined in Section~\ref{sec:smoother}}, respectively. We apply the smoothing operator $M$ times for presmoothing and $N$ times for postsmoothing \rev{on each level}, which is called the V(M,N) cycle. While Athena++ supports $M,N=0,1,2$, we typically use the V(1,1) cycle.

The algorithm applies $MULTIGRID()$ recursively until it reaches the coarsest level. Note that $MULTIGRID()$ on the coarser levels operates not on the potential itself but on the defect. This is because the self-gravity Poisson equation is linear and therefore the defect equation eq.(\ref{eq:defecteq}) can be used instead of the Poisson equation eq.(\ref{eq:discretized3}). This algorithm is called the {\bf multigrid correction} scheme.

Starting from an initial guess, we repeat this iteration procedure until a certain convergence threshold is satisfied, as discussed in Section~\ref{sec:criterion}. The schematic picture of the V-cycle multigrid iteration is shown in Figure~\ref{fig:cycles} (a).

\subsubsection{Full Multigrid}\label{sec:fmg}
The full multigrid method \citep{fmg} is a more sophisticated algorithm to solve the system. In this method, we start from the coarsest level where the solution can be easily calculated. The coarse-grid solution is prolongated to the finer level, and is used as an initial condition for the multigrid iteration to improve the fine-grid solution. By repeating this procedure, we can obtain a reasonably good solution once we reach the finest level. The pseudocode of this algorithm is as follows:\\
{\bf Algorithm 2:} Full multigrid
\begin{algorithmic}
\Procedure{$FMG$}{$L_h,\bm{f}_h$}
\State 1. Restrict source to coarser levels:
\State \hspace{3ex}Repeat $\bm{f}_{l-1} = I_{l}^{l-1}\bm{f}_{l}$ for all the levels.
\State 2. Solve the coarsest level: \State \hspace{3ex}$\bm{\phi}_0^n=COARSEST(\bm{0},L_0,\bm{f}_0^n)$.
\For{$l\leq$ finest}
  \State 3. Prolongate solution: $\bm{\phi}_l^n = {I'}_{l-1}^{l}\bm{\phi}_{l-1}^n$.
  \State 4. Apply Multigrid: \State \hspace{3ex}$\bm{\phi}_l^n = MULTIGRID(\bm{\phi}_{l}^n,L_{l},\bm{f}_{l},l)$.
\EndFor
\EndProcedure
\end{algorithmic}
$MULTIGRID()$ used here is the same as in Algorithm 1. The prolongation of the solution \rev{in the full multigrid method}, or the FMG prolongation, requires a higher order prolongation operator ${I'}_{l-1}^{l}$, which is explained in \ref{sec:operators}. The schematic picture of the full multigrid method with the V(1,1) cycles is shown in Figure~\ref{fig:cycles} (b).

It is remarkable that the full multigrid method does not require any initial guess, compared to the multigrid iteration. Instead, this algorithm fully utilizes the geometric structure to construct the initial guess for the finer levels. While we may apply $MULTIGRID()$ more than once per level, practically one application per level is sufficient to achieve the true error of the order of the truncation error compared to an analytic solution after the full multigrid algorithm. However, the defect and the discretized error remain considerably large because the numerical solution still contains error and noises, and we need to apply additional V-cycle iterations to obtain a fully-converged solution. We discuss the error behaviors in Section~\ref{sec:tests}.

\subsubsection{Full Approximation Scheme} \label{sec:fas}
In the multigrid correction scheme presented above, the defect equation is solved when the multigrid algorithm is applied on the coarser levels. This is efficient because we need to restrict only the defect, \rev{and not} the potential. This is possible because the Poisson equation and the defect equation are linear and have the same shape as in eqs (\ref{eq:discretized3}) and (\ref{eq:defect}). However, for non-linear problems, this is not satisfied. Also, when we apply the multigrid methods on the AMR hierarchy, this is not applicable because the defect is defined for a certain resolution and mathematical operations between the defects defined on different AMR levels are ill-defined.

When mesh refinement is in use (and for future extension to non-linear physical processes), we use the Full Approximation Scheme or Full Approximation Storage \citep[FAS,][]{fmg}. In FAS, we restrict the potential itself in addition to the defect, and solve the original equation (\ref{eq:discretized3}) on every multigrid level. This is achieved using the following algorithm:\\
{\bf Algorithm 3:} Full Approximation Scheme Iteration
\begin{algorithmic}
\Procedure{$FAS$}{$\bm{\phi}_{l}^n,L_{l},\bm{f}_{l},l$}
\State 1. Presmooth: $\bm{\bar{\phi}}_{l}^n=SMOOTH^M(\bm{\phi}_{l}^n,L_{l},\bm{f}_{l})$.
\State 2. Compute defect: $\bm{d}_{l}^n = \bm{f}_{l}-L_{l}\bm{\bar{\phi}}_{l}^n$.
\State 3. Restrict defect: $\bm{d}_{l-1}^n=I_{l}^{l-1}\bm{d}_{l}^n$.
\State 4. Restrict potential: $\bm{\bar{\phi}}_{l-1}^n=I_{l}^{l-1}\bm{\bar{\phi}}_{l}^n$.
\State 5. Compute source term: $\bm{f}_{l-1}^n=\bm{d}_{l-1}^n+L_{l-1}\bm{\bar{\phi}}_{l-1}^n$.
\State 6. Calculate coarse-grid solution:
\If{$l = 1$}
  \State 6.1 Solve the coarsest level: \State \hspace{4ex}$\hat{\bm{\psi}}_0^n=COARSEST(\bm{\bar{\phi}}_{l-1}^n,L_0,\bm{f}_0^n)$.
\Else
  \State 6.2 Apply $FAS$ recursively: \State \hspace{4ex}$\hat{\bm{\psi}}_{l-1}^n=FAS(\bm{\bar{\phi}}_{l-1}^n,L_{l-1},\bm{f}_{l-1}^n,l-1)$.
\EndIf
\State 7. Compute coarse-grid correction: \State \hspace{3ex}$\hat{\bm{\delta}}_{l-1}^n = \hat{\bm{\psi}}_{l-1}^n - \bar{\bm{\phi}}_{l-1}^n$.
\State 8. Prolongate correction: $\hat{\bm{\delta}}_{l}^n=I_{l-1}^{l}\hat{\bm{\delta}}_{l-1}^n$.
\State 9. Apply correction: ${\bm{\phi}'}_{l}^n=\bm{\bar{\phi}}_{l}^n+\hat{\bm{\delta}}_{l}^n$.
\State 10. Postsmooth: $\bm{\phi}_{l}^{n+1}=SMOOTH^N({\bm{\phi}'}_{l}^n,L_{l},\bm{f}_{l})$.
\State {\bf return} $\bm{\phi}_{l}^{n+1}$.
\EndProcedure
\end{algorithmic}
This scheme is equivalent to the multigrid correction scheme for a linear system on uniform grid. Compared to the multigrid correction scheme, this scheme requires additional computations for restriction of the potential, calculation of the RHS, and calculation of coarse-grid correction. However, the restriction is relatively less expensive, and the other additional computations are needed only on coarser levels that have much fewer cells. Therefore, additional cost from FAS is moderate. The FAS version of the full multigrid method is obtained by replacing $MULTIGRID()$ with $FAS()$ in Algorithm 2.

\subsubsection{Boundary Conditions}\label{sec:boundary}
Boundary conditions are required for the smoothing operations and prolongation operations on every level of the multigrid hierarchy. The multigrid methods are compatible with various boundary conditions, and we have implemented periodic, zero-gradient, zero-fixed boundary conditions as well as isolated boundary conditions using a multipole expansion in Athena++. An interface to enroll user-defined boundary conditions is also provided. The implementation of the multipole expansion is described in Appendix~\ref{sec:multipole}.

For the multigrid correction method, specified physical boundary conditions are applied on the finest level and zero-fixed boundary conditions are used for the defect equation on the coarser levels unless periodic boundary conditions are specified. When FAS is in use, we need to apply physical boundary conditions at every level consistently.

In Athena++, the boundary conditions are set in the ghost cells \rev{(whereas the other cells that are updated by the solvers are called ``active cells")}. For periodic and zero-gradient boundary conditions, the boundary values are set at cell-centers of the ghost cells. When other (e.g., isolated or fixed) boundary conditions are specified, we set the ghost cell values so that the value on the surface of the computing domain match the specified boundary values. This is because the cell-center positions of the ghost cells shift between levels.

Because the potential appears only within the Laplacian operator in the Poisson equation (\ref{eq:poisson}), the zero point of the potential is arbitrary. If at least one of the boundary conditions contains a fixed value, it sets the zero point. For the multipole expansion, it is natural to set it to zero at infinity. If all the boundary conditions do not set any fixed value (e.g., all the boundaries are periodic or zero-gradient), a Poisson solver cannot find the zero point by itself. In this case, we use the deviations from the averages of the source function and the potential, $\delta f=f-\langle f\rangle$ and $\delta \phi = \phi-\langle \phi\rangle$ where $\langle u \rangle$ is the volume average of $u$. The Poisson equation retains the same shape as $\nabla^2\delta\phi=\delta f$, but now we set the potential zero point to zero so that the average of the potential is $\langle \delta \phi\rangle=0$.

\subsubsection{Convergence Criterion}\label{sec:criterion}
In Athena++, two iteration modes are available. In the ``FMG" (Full MultiGrid) mode, the full multigrid method is used first, and then V-cycle iterations are applied to improve the solution. In the ``MGI" (MultiGrid Iteration) mode, one has to provide an initial guess and V-cycle iterations are applied repeatedly. In practical hydrodynamic simulations, one can use the potential from the previous timestep as the initial guess, but usually it is not as good as the numerical solution after one sweep of the full multigrid method. In general, we recommend the FMG mode because it requires no initial guess and performs better in most cases \rev{as shown in Section~\ref{sec:tests}}.

In both modes, the V-cycle iteration is repeated until a certain convergence threshold is met or until convergence saturates and the error reduction stalls. As the discretized error (\ref{eq:error}) and true error (\ref{eq:trueerror}) are unknown beforehand, we use the defect (\ref{eq:defect}) as the convergence threshold. Specifically, we use the volume-averaged root-mean-square of the defect :
\begin{eqnarray}
d^n &=& \sqrt{\frac{\sum_{i,j,k}{|{d}^n_{i,j,k}|^2} \Delta V_{i,j,k}}{\sum_{i,j,k} \Delta V_{i,j,k}}},\label{eq:threshold}\\
{d}^n_{i,j,k}&=&4\pi G \rho_{i,j,k}-\left(\phi^n_{i-1,j,k}+\phi^n_{i+1,j,k}+\phi^n_{i,j-1,k}\right.\nonumber\\&+&\left.\phi^n_{i,j+1,k}+\phi^n_{i,j,k-1}+\phi^n_{i,j,k+1}-6\phi^n_{i,j,k}\right)/h^2,\hspace{2em}
\end{eqnarray}
where $\Delta V_{i,j,k}$ is the volume of each cell and the summation is over all the cells in the computing domain.
Here, the defect is calculated on the finest (unrestricted) level. Once this value gets smaller than a user-specified threshold, the code stops. A user has to carefully specify a resonably-small threshold value for this, or set it to zero to repeat the iteration until convergence saturates. Alternatively, one can use a fixed number of iterations if it is possible to determine the number of iterations needed to achieve demanded accuracy. \rev{Because the demanded accuracy and corresponding defect threshold depend on problems, there is no threshold value that can be used universally. We recommend users to try a few different parameters to determine the convergence threshold before starting a production run.}

\section{Numerical Implementation}
\rev{In this section, we first review relevant features of the Athena++ code briefly (Section \ref{sec:athenapp}). Then we present our implementations of the multigrid solvers on uniform grids (Section \ref{sec:uniformimp}) and with AMR (Section \ref{sec:amrimp}). Lastly, we discuss the cost of the multigrid solvers in Section \ref{sec:mgcost}.}

\subsection{Brief Review of Athena++} \label{sec:athenapp}
Athena++ \citep{athenapp} is a public AMR framework supporting various physical processes and features for astrophysical applications, including non-uniform mesh spacing, Cartesian and curvilinear coordinates, static and adaptive mesh refinement, magnetohydrodynamics (MHD), special and general relativity \citep{athenappgr}, chemical reactions, general equation-of-states \citep{athenappeos}, particles (Yang et al. in prep.), shearing box and orbital advection (Ono et al. in prep.), non-ideal MHD effects and other diffusion processes, as well as other physics. Beyond the public version, various extensions such as cosmic-ray transport \citep{athenappcr}, time-dependent radiation transport \citep{athenapprad}, post-processing general relativistic radiation transport \citep{blacklight} and full general relativity \citep{grathenapp}, are being actively developed in the community as well. For self-gravity, there is already an implementation based on FFTs for uniform grids (developed by C.-G. Kim). The code is publicly available on GitHub, and documentation and a tutorial are available on the website\footnote{\url{https://www.athena-astro.app/}}.

Athena++ is parallelized using the standard domain decomposition technique using MPI and OpenMP. The computing domain is split into units called \texttt{MeshBlocks}, each of the same logical size (i.e., containing the same number of cells). In Athena++, MPI processes and OpenMP threads are almost equivalent; each process or thread owns one or more \texttt{MeshBlocks}. Each \texttt{MeshBlock} has additional ghost cells at each boundaries for communication between \texttt{MeshBlocks} and also for physical boundary conditions. The ghost cell data are transferred via MPI messages or memory copy depending on they are on the same MPI process or not.

Athena++ adopts an octree block-based AMR design. In this design, a \texttt{MeshBlock} is not only the domain decomposition unit for parallelization but also the unit for mesh refinement. When a \texttt{MeshBlock} is refined, it is split into eight finer \texttt{MeshBlocks} in 3D. Neighboring \texttt{MeshBlocks} are also refined so that only \texttt{MeshBlocks} on the same level or one level different contact each other. This design is computationally efficient, because all the \texttt{MeshBlocks} are logically uniform grids of the same shape, and highly-optimized codes for uniform grids can be applied. The design also works well for parallelization because only a few simple patterns for communication between \texttt{MeshBlocks} need to be considered. We use the Z-ordering to assign globally unique IDs to \texttt{MeshBlocks}, and distribute them as evenly as possible for parallelization. 

One of unique designs of Athena++ is dynamical scheduling using a \texttt{TaskList}. The simulation program is split into small \texttt{Tasks} with associated dependencies, which are then assembled into a \texttt{TaskList}. Normally \texttt{Tasks} are processed sequentially from the beginning of a \texttt{TaskList}, but when the code has to wait for communication associated with a particular \texttt{Task} to arrive, it automatically processes the next \texttt{Tasks} in the \texttt{TaskList} that do not depend on the communication. This enables overlapping of computation and communication automatically. In addition, as this design separates logic within each physics module and relation between modules, it improves flexibility and modularity of the design.

Athena++ uses uniform timestepping even when AMR is in use. This means that all \texttt{MeshBlocks} on all levels share the same timestep.  While this methods does not minimize the number of operations, it often is the most efficient because it does not require complicated scheduling and load-balancing logic nor synchronization between AMR levels, and parallelization is straightforward. Moreover, non-local physics such as self-gravity and radiation transport with an implicit time integrator are captured more accurately since the solution at every level is at the same time and temporal interpolation is not necessary. Thus, the uniform timestepping as adopted in Athena++ is advantageous for applications involving non-local physics. For complete details of Athena++, we refer readers to the method paper \citep{athenapp}. \rev{In particular, the design of the AMR framework is discussed in Section 2.1 and the TaskList implementation is in Section 2.5.}

\subsection{Uniform Grid}\label{sec:uniformimp}
We first describe our implementation of the multigrid methods on uniform grids in this section. \rev{Our} implementation does not depend on any external library (except for MPI), and the code is written in C++11.

\subsubsection{Multigrid Hierarchy and Data Structure}
We decompose the computational domain using the same \texttt{MeshBlocks} as in the main (hydrodynamics or MHD) part. Because we apply the restriction and prolongation operators on each \texttt{MeshBlock}, each \texttt{MeshBlock} has to be logically cubic and the number of active cells in each direction must be power of two, i.e., $(n_x, n_y, n_z) = (2^n, 2^n, 2^n)$. This is more restrictive than the AMR of Athena++. However, there is no limitation in the number of \texttt{MeshBlocks} in each direction. We define the coarsest level where the grid cannot be restricted anymore as level 0. Let $(N_{l,x}, N_{l,y}, N_{l,z})$ correspond to the numbers of cells in each direction on level $l$. If the number of cells on level 0 is $(N_{0,x}, N_{0,y}, N_{0,z})$, then $(N_{l,x}, N_{l,y}, N_{l,z}) = (2^lN_{0,x}, 2^lN_{0,y}, 2^lN_{0,z})$. There is a certain level $m$ where the number of the cells matches the number of the \texttt{MeshBlocks} $(N_{m,x}, N_{m,y}, N_{m,z})=(N_{{\rm MB},x}, N_{{\rm MB},y}, N_{{\rm MB},z})$. In total there are $L=m+n$ levels, and the total number of cells in each direction is  $(N_{x}, N_{y}, N_{z})=(2^nN_{{\rm MB},x},2^nN_{{\rm MB},y},2^nN_{{\rm MB},z})=(2^{m+n}N_{0,x},2^{m+n}N_{0,y},2^{m+n}N_{0,z})$, where $N_{{\rm MB}}$ is the number of \texttt{MeshBlocks} in each direction.

The potential, source function and defect on levels from $m$ to $m+n$ are stored as multi-dimensional arrays in \texttt{Multigrid} class in each \texttt{MeshBlock}. On the other hand, those data on levels from $0$ to $m$ are stored in a separate \texttt{Multigrid} class, which we call \texttt{RootGrids}. In addition to the active cells, ghost cells are allocated on each side of the \texttt{Multigrids} for communication with neighboring \texttt{Multigrids}. For self-gravity, we need only one layer of the ghost cells, $N_G=1$. When FAS is in use, additional storage is allocated for the restricted potential on coarser levels. The data on level $m$ are stored both in \texttt{RootGrids} and \texttt{Multigrid} for communication and synchronization between these objects. Note that each \texttt{Multigrid} object contains only $1\times1\times1$ cell on level $m$. Hereafter, we refer levels $0$ to $m$ as \texttt{RootGrid} levels and $m+1$ to $m+n$ as \texttt{MeshBlock} levels.

\subsubsection{V-Cycle Algorithm on Uniform Grid}
In the V-cycle algorithm, we start the iteration procedure from the finest level, whose data are stored in \texttt{Multigrids}. To start up the algorithm, the source function $\bm{f}$ is calculated using the mass density loaded from the \texttt{MeshBlocks}, and the initial guess $\bm{\phi}^0$ is loaded into the potential data.

As in Algorithm 1, each iteration starts with pre-smoothing using the RBGS smoother. We then calculate the defect $\bm{d}$, restrict and transfer it to the coarser level. This procedure is repeated recursively until it reaches level $m$, where each MeshBlock is restricted to $1\times1\times1$ cell and cannot be restricted anymore. At this point, the data on level $m$ are collected from all the \texttt{Multigrids} and transferred to the array on level $m$ in \texttt{RootGrids}. Then, the recursion continues to the \texttt{RootGrid} levels until it reaches the coarsest level $0$. \rev{The coarsest level solution is easily obtained using the coarsest-level solver described in Section\ref{sec:smoother} because there are only a few degrees of freedom.}

Then, the correction on the coarse level is prolongated into the finer level. After adding the correction, we apply the RBGS smoother for post-smoothing. This procedure is repeated until it reaches level $m$ where the data are transferred from \texttt{RootGrids} to \texttt{Multigrids}, then continues to the finest level $m+n$. This completes one iteration of the V-cycle algorithm.

After each V-cycle iteration, the average defect (\ref{eq:threshold}) is computed and compared with a given convergence threshold. Once the convergence criterion discussed in Section~\ref{sec:criterion} is satisfied, the algorithm is completed and returns the potential as the result. The potential is then used in the hydrodynamics or MHD part to calculate the gravitational acceleration.

The logic remains almost the same even with FAS enabled. In this case, the potential is also restricted along with the defect, and additional computations of the RHS and coarse-grid correction are performed accordingly.

\subsubsection{Time Integration of Hydrodynamic Source Terms}
The time integration of the hydrodynamics or MHD part is done in the main integrator of Athena++, which is either the second-order van-Leer, second-order Runge-Kutta or third-order Runge-Kutta. The integration procedure is processed using \texttt{TaskList}, but the multigrid solver is implemented using a separate \texttt{TaskList} from the main integrator. In every substep, we load the density given from the time integrator to the multigrid module, run the multigrid solver, pass the potential back to the hydrodynamics/MHD module, and run the main integrator. This is not the most cost-effective implementation as it requires a multigrid solve for each substep of the integrator, but it is simple and compatible with any time integrator.

The momentum and energy source terms by the gravitational force are calculated using the cell-centered source term scheme \citep{mullen2021}. For example, the momentum source term in the $x$-direction is
\begin{eqnarray}
g_{x,i-1/2,j,k} & = & -\frac{\phi_{i,j,k}-\phi_{i-1,j,k}}{h}, \nonumber\\
g_{x,i+1/2,j,k} & = & -\frac{\phi_{i+1,j,k}-\phi_{i,j,k}}{h}, \nonumber\\
g_{x,i,j,k} & = & \frac{g_{x,i-1/2,j,k}+g_{x,i+1/2,j,k}}{2}, \nonumber\\
(\rho \bm{g})_{x,i,j,k}&=&\rho_{i,j,k}g_{x,i,j,k}, \nonumber
\end{eqnarray}
and the energy source term is
\begin{eqnarray}
&&(\rho \bm{v}\cdot\bm{g})_{i,j,k}\nonumber\\
&=&\frac{1}{2}\left[(\rho \bm{v})_{x,i-1/2,j,k}g_{x,i-1/2,j,k}+(\rho \bm{v})_{x,i+1/2,j,k}g_{x,i+1/2,j,k}\right]\nonumber\\
&+&\frac{1}{2}\left[(\rho \bm{v})_{y,i,j-1/2,k}g_{y,i,j-1/2,k}+(\rho \bm{v})_{y,i,j+1/2,k}g_{y,i,j+1/2,k}\right]\nonumber\\
&+&\frac{1}{2}\left[(\rho \bm{v})_{z,i,j,k-1/2}g_{z,i,j,k-1/2}+(\rho \bm{v})_{z,i,j,k+1/2}g_{z,i,j,k+1/2}\right],\nonumber\\
& &
\end{eqnarray}
where $(\rho \bm{v})_{x,i+1/2,j,k}$, etc. is the mass flux at the cell interface returned from a Riemann Solver. This implementation is curl-free, contrary to the implementation using the gravitational flux \citep[e.g.]{jiang2013}\footnote{A curl-free implementation based on the \rev{gravitational} flux is proposed in \citet{mullen2021}}. On the other hand, for time integration, we simply use one of the standard integrators in Athena++ and we do not use the conservative scheme for the energy term proposed in \citet{mullen2021}. Therefore, our scheme does not exactly conserve the energy, only to the accuracy of the scheme. It should be noted that both momentum and energy are not conserved exactly to the round-off error because there is non-zero residual in the potential obtained with the multigrid solver in general.

Technically, it is possible to combine the self-gravity \texttt{TaskList} and hydrodynamics/MHD \texttt{TaskList} and maximize communication-computation overlapping. The reason why we implement the self-gravity as a separate \texttt{TaskList} from the main integrator is partly for simplicity, but also for future extension to heterogeneous parallelization, in which some nodes process self-gravity while other nodes work on other physical processes. Because the multigrid solver is less scalable than the MHD solver as shown later in \rev{Section} \ref{sec:tests}, such an advanced parallelization technique could be useful in the future to improve the overall parallelization efficiency.

\subsubsection{FMG Algorithm on Uniform Grid}
The implementation of FMG is straightforward on top of the V-cycle algorithm. First, the source function is restricted from the finest levels to the coarsest levels and stored on each level. Then, we start the algorithm from the coarsest level. On the coarsest level, the solution is obtained either analytically or using a few sweeps of the iterative solver. Then, this solution is prolongated using the third-order tricubic interpolation to the finer level. Using the prolongated solution as the initial guess, the V-cycle algorithm is applied to improve the solution on this level. We repeat this procedure until it reaches the finest level, as shown in Figure~\ref{fig:cycles} (b). Once the solution on the finest level is obtained, we apply the V-cycle repeatedly to improve its accuracy until the convergence criterion is satisfied.

\subsubsection{Parallelization}
In order to run large-scale simulations on modern supercomputers, parallelization is of crucial importance. Because parallelization of FMG is trivial once the V-cycle is parallelized, we describe the parallelization of the V-cycle algorithm here.

As in the hydrodynamic/MHD part in Athena++, the multigrid solver is parallelized using both MPI and OpenMP. When parallelization is enabled, each MPI process (or OpenMP thread in case OpenMP or hybrid parallelization is in use) owns one or more \texttt{MeshBlocks} and associated \texttt{Multigrids}. The data on the \texttt{MeshBlock} levels are stored in \texttt{Multigrids} and distributed to each process. Parallelization in the \texttt{MeshBlock} levels is similar to the hydrodynamic/MHD part. The data on the boundaries are transferred to and from neighboring \texttt{Multigrids} through the ghost cells before the smoothing and prolongation operations. For the RBGS smoother, the boundaries are communicated before each sweep of red and black. Using \texttt{TaskList}, computations and communications are automatically overlapped.

Once the algorithm reaches the transition level $m$ where each \texttt{Multigrid} is reduced to a single cell, the data on this level are collected and transferred to \texttt{RootGrids} on every MPI rank using {\it MPI\_Allgather}\footnote{Logically, this implementation is the same as collecting the data into one MPI rank with {\it MPI\_Gather}, processing the \texttt{RootGrids}, then distributing the data back to \texttt{Multigrids} with {\it MPI\_Scatter}. However, we chose our implementation considering that one {\it MPI\_Allgather} requires less synchronization than the combination of {\it MPI\_Gather} and {\it MPI\_Scatter}.}. Then, every rank performs the multigrid algorithm on the \texttt{RootGrid} levels. Once it comes back to level $m$, then the data are transferred to each \texttt{Multigrid}. Then, again each MPI process or OpenMP thread calculates the assigned \texttt{Multigrids}.


\begin{figure*}[t]
\begin{center}
\includegraphics[width=\textwidth]{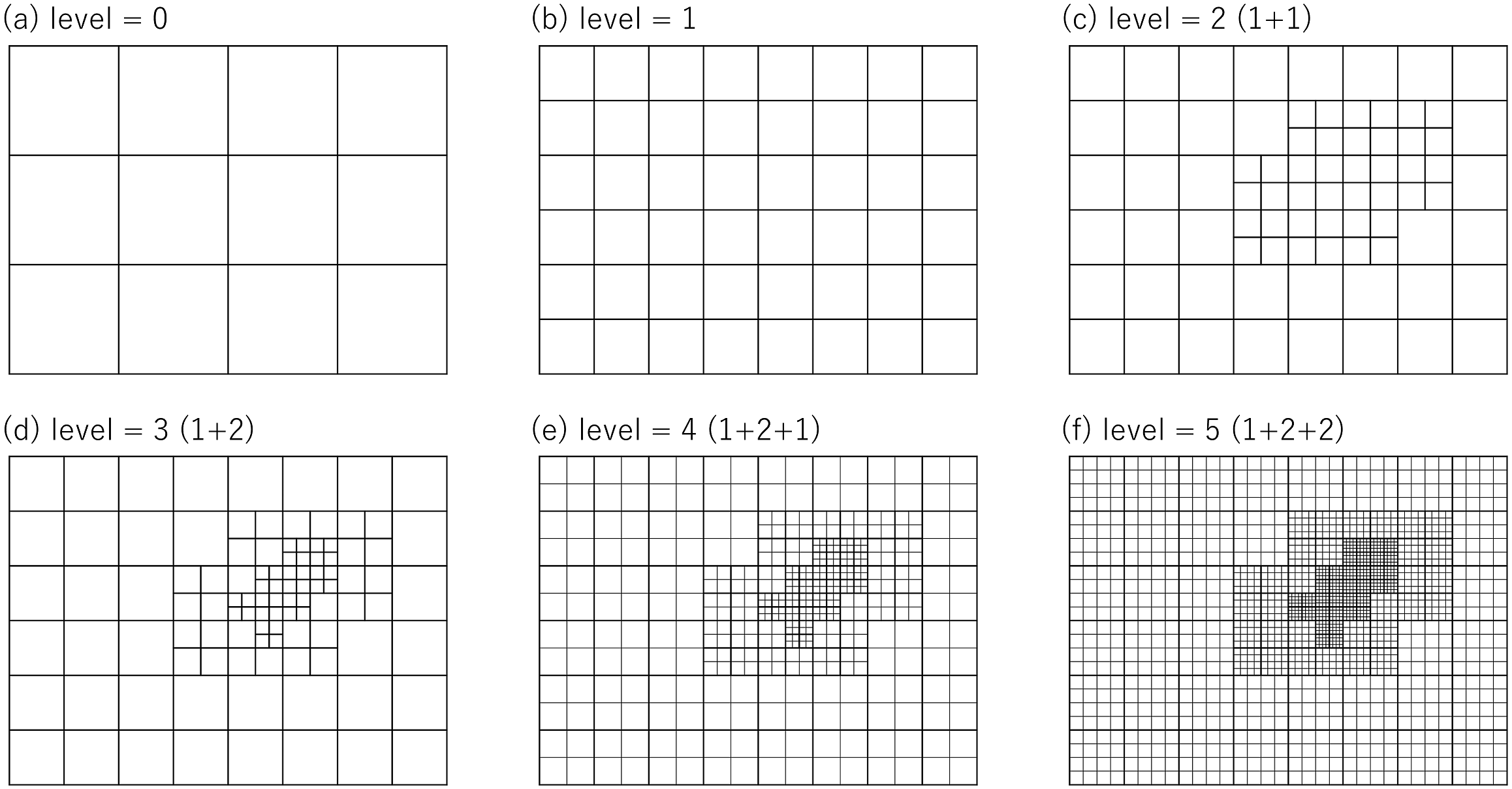}
\end{center}
\caption{An example of a multigrid level hierarchy with mesh refinement in 2D. Thick tiles represent \texttt{Multigrids} (associated with \texttt{MeshBlocks}) and \texttt{Octets}, while thin tiles are cells within each \texttt{Multigrid}. (a) Level $0$ consisting of $4\times 3$ cells. (b) Level $1$ with $8\times 6$ cells. (c) Level $2$ with the first level of \texttt{Octets}. (d) Level $3$ with the second refinement level of \texttt{Octets}. (e) Level $4$ where each \texttt{Multigrid} is refined to $2\times2$ cells. (f) Level $5$ with \texttt{Multigrids} consisting of $4\times4$ cells.}
\label{fig:amr}
\end{figure*}

\subsection{Mesh Refinement}\label{sec:amrimp}
We do not discuss generation, destruction and load-balancing of \texttt{MeshBlocks} here because these processes are unchanged by the addition of multigrid, and are controlled by teh AMR framework of Athena++. For details, see the method paper \citep{athenapp}.

There are, however, two major changes to the multigrid algorithm itself with mesh refinement. One is manipulation of the refined \texttt{Multigrids} on coarse levels, and the other is boundary communication between neighboring \texttt{Multigrids} on different levels.

\subsubsection{V-Cycle Algorithm on AMR Hierarchy}
An example of the multigrid operations on an AMR hierarchy is shown in Figure~\ref{fig:amr}. Compared to the uniform grid, there are additional levels where refined \texttt{MeshBlocks} are inserted as shown in Panels (c) and (d). We refer these levels as \texttt{Octet} levels. {Because of the octree AMR design of Athena++, a \texttt{MeshBlock} is always refined into a set of $2\times2\times2$ finer MeshBlocks in 3D ($2\times2$ in 2D). We call this set \texttt{Octet} in Athena++, and we perform the smoothing, restriction and prolongation operations on each \texttt{Octet}. Similar to \texttt{Multigrids}, the potential, source function and defect data are allocated in each \texttt{Octet} object, which contains $2\times2\times2$ cells and ghost cells on each side for boundary communications. The relations between the finer and coarser \texttt{MeshBlocks}/\texttt{Octets} as well as the connections between neighboring \texttt{MeshBlocks}/\texttt{Octets} are calculated using the \texttt{MeshBlockTree} facility in Athena++.

When AMR is enabled, FAS is used as discussed in Section~\ref{sec:fas}. The algorithm on the \texttt{MeshBlock} levels is not very different from the uniform grid case. Starting from the finest level, it goes up the multigrid levels toward the coarser levels (Figure~\ref{fig:amr} (f) to (e)). The level boundaries are described in the next section.

When it reaches the level where each \texttt{MeshBlock} is reduced to a single cell (Figure~\ref{fig:amr} (d)), the data on \texttt{Multigrids} that are not refined are directly transferred to the \texttt{RootGrids}, while the data on \texttt{Multigrds} generated by refinement are transferred to corresponding cells in \texttt{Octets}. Then we apply smoothing on the \texttt{Octets} on the finest level and restrict the them to the coarser levels (Figure~\ref{fig:amr} (d) to (b)). For this smoothing operation, the boundaries contacting other \texttt{Octets} on the same level are actively updated, while the boundaries facing the coarser level are given as fixed boundary values prolongated from the coarser level. Once all the \texttt{Octet} levels are removed by repeating this procedure, the data are transferred to the corresponding cells in the \texttt{RootGrids}. Then, the algorithm continues toward the coarsest level as in the uniform grid case (Figure~\ref{fig:amr} (b) to (a)). 

Once it reaches the coarsest level, the solution on the coarsest level is calculated in the same way as in the uniform grid case. Then the algorithm goes back to the finer levels by prolongation (Figure~\ref{fig:amr} (a) to (b)). When it reaches the first \texttt{Octet} level, the \texttt{Octets} on the first level are generated by prolongating the corresponding cells on the \texttt{RootGrids}, and then the smoothing operator is applied on the newly generated \texttt{Octets} (Figure~\ref{fig:amr} (c)). This procedure is repeated toward the finer \texttt{Octet} levels (Figure~\ref{fig:amr} (c) to (d)). When it comes back to the \texttt{MeshBlock} level, the data are transferred to \texttt{Multigrids}, and then they are prolongated and smoothed recursively until it reaches the finest level (Figure~\ref{fig:amr} (e) to (f)). This completes one V-cycle iteration on an AMR hierarchy.

We note that in some implementations of self-gravity solvers described in the literature \citep[e.g.][]{art,ramsesmg,enzo,dispatchgrav}, each refined level is solved separately using the coarser level data as boundary conditions. However, such a level-by-level or block-by-block approach is not self-consistent as the coarse level solutions do not incorporate small-scale structure on the finer levels, which then implies the finer level solutions are not accurate because the boundary conditions obtained from the coarser levels are not correct. In contrast, the iteration algorithm on AMR meshes used here solves the Poisson equation across all the refinement levels consistently. \rev{We present a rough comparison between these approaches in Appendix~\ref{sec:amr_lbl}.}

Since FMG is built on top of the V-cycle algorithm, its implementation is straightforward once the V-cycle with mesh refinement is implemented. Parallelization of the multigrid algorithm with mesh refinement is similar to the uniform grid case. 

Our iteration algorithm on AMR is slightly different from what is proposed in \citet{multigrid}. In the first part of the V-cycle algorithm, we restrict all the \texttt{Multigrids} until they are reduced to a single cell and then remove them. In contrast, with the method of \citet{multigrid}, locally-refined \texttt{Multigrids} are removed first. We chose this design because our implementation can maintain load balance better, since the number of \texttt{Multigrids} remains the same until it reaches the \texttt{Octet} levels.

\begin{figure*}[t]
\begin{center}
\includegraphics[width=\textwidth]{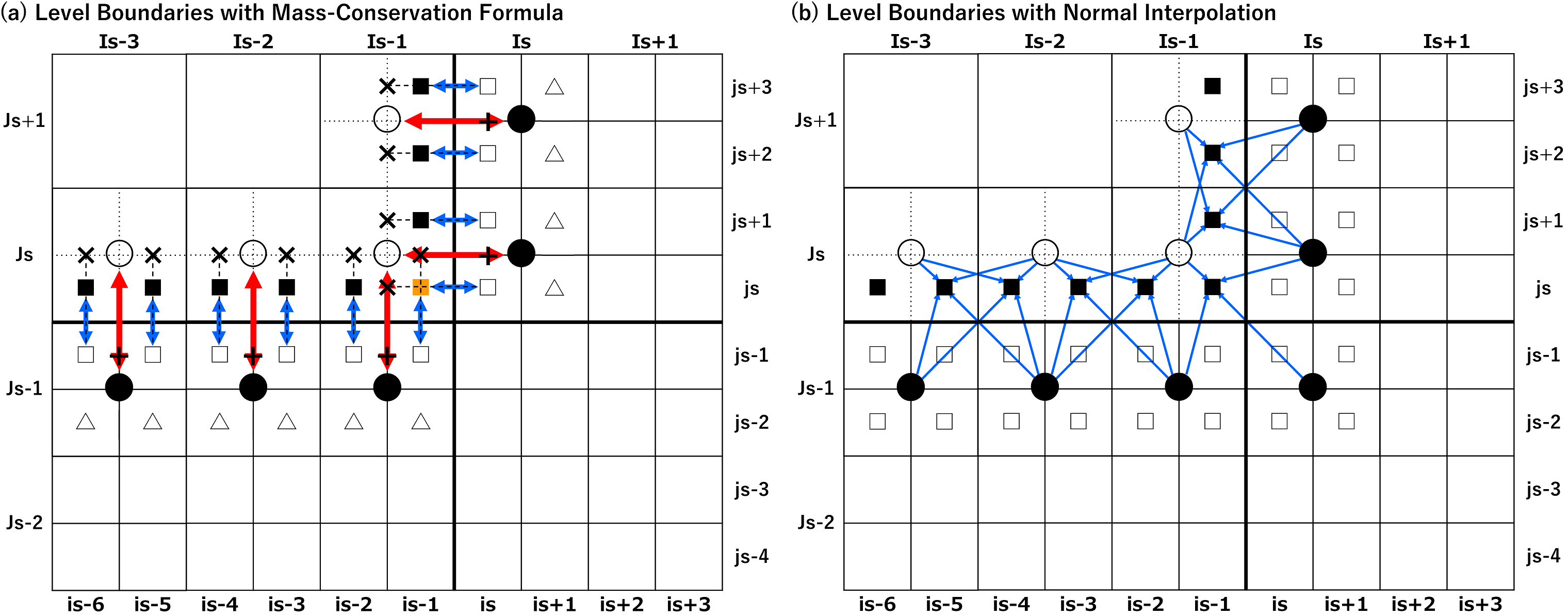}
\end{center}
\caption{2D illustration of the ghost cell interpolation methods using (a) the mass conservation formula and (b) normal interpolation. Thick lines indicate the boundaries between \texttt{Blocks}. The \texttt{Block} on the top left corner is on the coarser level, while the other \texttt{Blocks} are on the finer level. Thin solid lines indicate active cells , while thin dotted lines indicate ghost cells. Open/filled circles are active/ghost cells on the coarse level, while open/filled squares are active/ghost cells on the finer level. Open triangles are active cells on the finer level used only for calculating restricted potential. (a) The ghost cell values are calculated so that the area-averaged gradients across the level boundaries match using the mass conservation formula. (b) The ghost cell values are calculated using the trilinear prolongation and volume weighted restriction operators. See the text for detail.}
\label{fig:lb}
\end{figure*}

\subsubsection{Level Boundaries}
On the \texttt{MeshBlock} and \texttt{Octet} levels, \texttt{Multigrids} and \texttt{Octets} on different levels contact each other. In this section, we refer \texttt{MeshBlocks} and \texttt{Octets} as \texttt{Blocks} as the boundary operations are the same. It is of crucial importance for the AMR multigrid solver to consistently connect such level boundaries. In Athena++, this is done by properly computing the ghost cell values at level boundaries. For this purpose, we adopt a method similar to those proposed in \citet{masscons}. We use two types of level boundaries. One adopts a conservative formulation and is used for the boundary communications before smoothing operations and force calculation, while the other is a simpler ``normal" level boundary used before the prolongation operations. While the hydrodynamic/MHD part of Athena++ requires two or more layers of the ghost cells (depending on the spatial order of the reconstruction methods), the second-order multigrid solver implemented here uses only one ghost cell layer even with AMR, because the discretized Laplacian operator (\ref{eq:discretized2}) spans a seven-point stencil in 3D.

{\bf Conservative Formulation.}  In order for the Laplacian operator to be accurate at level boundaries, the \rev{gravitational acceleration} at cell interfaces at level boundaries must be the same for both the fine and coarse grids. For example, on the left surface in the $x$-direction, the following condition must be satisfied:
\begin{eqnarray}
\frac{\phi_{I_s,J,K}-\phi_{I_s-1,J,K}}{2h}\nonumber\\
=\frac{1}{4}\sum_{a=0}^1\sum_{b=0}^1 \frac{\phi_{i_s,j+a,k+b}-\phi_{i_s-1,j+a,k+b}}{h},\label{eq:masscons}
\end{eqnarray}
where $I_s$ and $i_s$ are the $x$-indexes of the first cells on the coarser and finer levels, $J,K$ and $j,k$ the $y$- and $z$-indexes of the corresponding cells on the coarser and finer levels, $h$  the resolution on the finer level. Note that this operation as well as the following computations should be replaced with area-weighted average or appropriately weighted interpolations if the grid spacing is not uniform and isotropic.  Note that the constraint expressed by equation~(\ref{eq:masscons}) enforces the truncation error in the calculation of the gradient to be divergence-free, which is equivalent to preventing an artificial or ``fake" mass at the boundary. Note that the AMR algorithms used to integrate the MHD equations also require flux-corrections at fine/coarse boundaries in order to conserve quantities such as mass and momentum, however with multigrid it is possible to adopt the appropriate difference formulae that satisfies conservation by its construction without any additional correction steps. 

As an example, let us consider a configuration illustrated in Figure~\ref{fig:lb} (a) and explain the level boundaries contacting in the $x$-direction at $i=i_s-1/2$ and $I=I_s-1/2$, but generalization to other configuration is straightforward. Here after the cell indices on the coarser and finer levels are represented using the capital and lowercase letters, and without any specific description, a coarse cell at $I,J,K$ corresponds to finer cells with indices $i,j,k$ and adjacent cells. 

First, we restrict the potential in the \rev{first} active cells on the finer level facing the coarser neighbor (open squares) using the area weighted average,
\begin{eqnarray}
\phi_{i_s,J,K}=\frac{1}{4}\sum_{a=0}^1\sum_{b=0}^1\phi_{i_s,j+a,k+b},\nonumber
\end{eqnarray}
 and send them to the coarser neighbor (black plus symbols). At the same time, we send the potential in the first active cells on the coarser level to the finer neighbor. The first active cells from the neighbors contacting on edges and corners are also sent to the finer neighbor as we need them later to calculate the transverse gradients. If these edge or corner neighbors are on the same level as the finer neighbor, we send restricted values as well (e.g. the filled circle at $I_s-1,J_s-1$ obtained by restricting the adjacent open squares and triangles).
 
Once the communications are completed, on the process which owns the finer \texttt{Block}, we calculate the transverse gradients on the coarser level:
\begin{eqnarray}
\begin{split}
D_{y,I_s-1,J,K}&=&\frac{\phi_{I_s-1,J+1,K}-\phi_{I_s-1,J-1,K}}{2h},\\
D_{z,I_s-1,J,K}&=&\frac{\phi_{I_s-1,J,K+1}-\phi_{I_s-1,J,K-1}}{2h}.
\end{split}\label{eq:transgrad}
\end{eqnarray}
Using these gradients, we interpolate the potential to the points aligned to the cell centers of the finer level (black crosses):
\begin{eqnarray}
\begin{split}
\phi_{I_s-1,j,k}&=&\phi_{I_s-1,J,K}-\frac{h}{2}D_{y,I_s-1,J,K}-\frac{h}{2}D_{z,I_s-1,J,K},\\
\phi_{I_s-1,j+1,k}&=&\phi_{I_s-1,J,K}+\frac{h}{2}D_{y,I_s-1,J,K}-\frac{h}{2}D_{z,I_s-1,J,K},\\
\phi_{I_s-1,j,k+1}&=&\phi_{I_s-1,J,K}-\frac{h}{2}D_{y,I_s-1,J,K}+\frac{h}{2}D_{z,I_s-1,J,K},\\
\phi_{I_s-1,j+1,k+1}&=&\phi_{I_s-1,J,K}+\frac{h}{2}D_{y,I_s-1,J,K}+\frac{h}{2}D_{z,I_s-1,J,K}.
\end{split}\label{eq:piint}
\end{eqnarray}
Note that the average potential of these four points is equal to the coarse potential $\phi_{I_s-1,J,K}$. Then, by interpolation between these points (black crosses) and the first active cells (white squares), we calculate the ghost cell values on the finer \texttt{Block} (filled squares)
\begin{eqnarray}
\phi_{i_s-1,j+a,k+b}=\frac{1}{3}\phi_{i_s,j+a,k+b}+\frac{\rev{2}}{3}\phi_{I_s-1,j+a,k+b},\label{eq:lbint}
\end{eqnarray}
where $a$ and $b$ are 0 or 1. At the same time, on the process which owns the coarser \texttt{Block}, we calculate the ghost values at the cell centers (filled circles) by extrapolating the gradient between the active cell centers (open circles) and the points received from the finer \texttt{Block} (black plus symbols):
\begin{eqnarray}
\phi_{I_s,J,K}=\frac{4}{3}\phi_{i_s,J,K}-\frac{1}{3}\phi_{I_s-1,J,K}.\label{eq:lbext}
\end{eqnarray}
The gradients used in the interpolation on the finer level (\ref{eq:lbint}) are calculated between the black crosses and open squares. On the other hand, the gradients used in the extrapolation on the coarser level (\ref{eq:lbext}) are calculated between the open circles and black plus symbols. Across each cell interface, the average of the gradients on the finer level matches the gradient on the coarser level. Therefore, these ghost values satisfy the mass conservation condition (\ref{eq:masscons}). In addition, this scheme requires minimum communications, only one set of the communications which are mutually independent.

When a coarser cell is facing more than one finer \texttt{Blocks}, prolongated ghost cells contacting the finer neighbors have multiple values for each finer neighbor. Such cells are called double (or triple if it is on the corner facing three finer neighbors) ghost cells (the orange square in Figure~\ref{fig:lb}). They do not cause any problem for the smoothing operations and force calculations, but we need a different method for boundary conditions for the prolongation operation, as the prolongation requires a consistent and unique value even in double ghost cells.

Although we use the same RBGS smoother even with AMR, it should be noted that the red-and-black pattern cannot be satisfied at level boundaries because one coarse cell faces four finer cells, which include two red and two black cells. We simply update all the ghost cells at the level boundaries after each smoothing operation.

\citet{masscons} used a different method in order to calculate the ghost cell values of the finer \texttt{Block} (filled squares). In their $\triangle$ stencil (see Figure~4 in \citet{masscons}), extrapolation of the transverse gradients of the first active cells (open squares) in the finer \texttt{Block} is used instead of the interpolation in eq.(\ref{eq:transgrad}). In contrast, our scheme (\ref{eq:transgrad}-\ref{eq:lbint}) corresponds to the $\Pi$ stencil in their paper. We find some cases where the $\Pi$ stencil behaves better, because the extrapolated transverse gradients used in the $\triangle$ stencil can be considerably different from actual ones.

{\bf Normal Interpolation.} For the prolongation operations including the higher order interpolation in the FMG algorithm, we adopt a different boundary condition. This is because we need to fill the ghost cells not only from the face neighbors but also the edge and corner neighbors, as the prolongation requires a larger stencil ($3\times3\times3$ cells) than the Laplacian operator (Figure~\ref{fig:stencil}). For this purpose, we simply calculate the ghost cells at the level boundaries using the same restriction and prolongation operations in the multigrid algorithm itself because the boundary conditions for the prolongation do not require the conservative formulation described above.

The interpolation procedure is illustrated in Figure~\ref{fig:lb} (b). For the ghost cells on the coarser \texttt{Blocks}, the active cells facing the level boundaries (open squares) on the finer level are restricted using (\ref{eq:restriction}), then sent to the ghost cells in the coarser \texttt{Blocks} (filled circles). At the same time, the active cells on the coarser \texttt{Blocks} facing the finer level (open circles) are sent to the finer \texttt{Blocks}. If some of the neighbors needed for the prolongation are on the finer level, those data are also restricted (filled circles). These data on the coarser level are prolongated using the trilinear prolongation operator (\ref{eq:trilinear}, blue arrows) into the ghost cells in the finer \texttt{Blocks} (filled squares). Again, these boundary communications are designed so that one set of the mutually-independent communications are required. Note that we use the trilinear prolongation even for the boundary conditions for the FMG prolongation because the trilinear prolongation does not need additional ghost cells, and also because we still can maintain the overall second-order accuracy with it.

\subsection{Cost of Multigrid}\label{sec:mgcost}
The majority of the computing cost of the multigrid methods come from the smoothing, restriction and prolongation operations, which are proportional to the total number of the cells. As the coarser level contains one-eighth cells of the finer level, the total number of the operations of one V-cycle is proportional to $N_x^3\left[1+\frac{1}{8}+\left(\frac{1}{8}\right)^2+\cdots\right]\sim \frac{8}{7}N_x^3$. Therefore, the computational complexity of our multigrid algorithms per sweep is $O(N_x^3)$. As the multigrid methods reduce the errors at all the wavelengths coherently, the number of iterations required to achieve a given accuracy does not depend on the number of the cells. Thus, the total computational cost of the multigird algorithms is also $O(N_x^3)$ or $O(N_{\rm tot})$. This is the advantage of the multigrid algorithms compared to FFT which is $O(N_{\rm tot}\log N_{\rm tot})$, and to classical iterative solvers such as SOR which is $O(N_x^4)$.

In order to perform very large simulations, the parallel scalability of the algorithm is of great importance and interest. In particular, we here focus on weak scaling, where we increase the size of the simulation proportional to the number of parallel processes and keep the load per each process constant. As the computational complexity of the multigrid algorithms is $O(N_{\rm tot})$, the multigrid methods are expected to achieve good parallel scaling as long as the overhead of communications is not significant and cost of the \texttt{RootGrids} and \texttt{Octet} levels is subdominant, as these parts are not fully parallelized in our implementation. The computational cost of the \texttt{RootGrid} and \texttt{Octet}levels should be relatively small because the degrees of freedom are only as large as the number of the \texttt{Multigrids}. However, for massively parallel simulations particularly with AMR, the \texttt{RootGrids} can have significant computational load and it may degrade the parallel performance if only one CPU core is used there. An additional layer of parallelization may be needed to improve the parallel performance in such cases, but we leave it for future work.

On uniform grids, the cost of the \texttt{RootGrids} is proportional to the number of the \texttt{MeshBlocks}, or $O(N_{{\rm MB},x}N_{{\rm MB},y}N_{{\rm MB},z})\sim O(N_{\rm MB}^3)$, while the cost of the \texttt{MeshBlocks} levels is $O(n^3)$ where $n$ is the number of the cells per \texttt{MeshBlock} in each direction. Therefore, it is favorable for parallel performance to use larger \texttt{MeshBlocks} on uniform grids. Roughly speaking, it is difficult to achieve good parallel performance if $N_{\rm MB}$ is larger than $n$. For AMR, the best \texttt{MeshBlock} configuration depends on problems and the size of \texttt{MeshBlocks} should be carefully selected so that an optimal balance between the performance and grid flexibility is achieved. We demonstrate the parallel performance of our multigrid solver in Section~\ref{sec:tests}.

\begin{figure*}[t]
\begin{center}
\includegraphics[width=\textwidth]{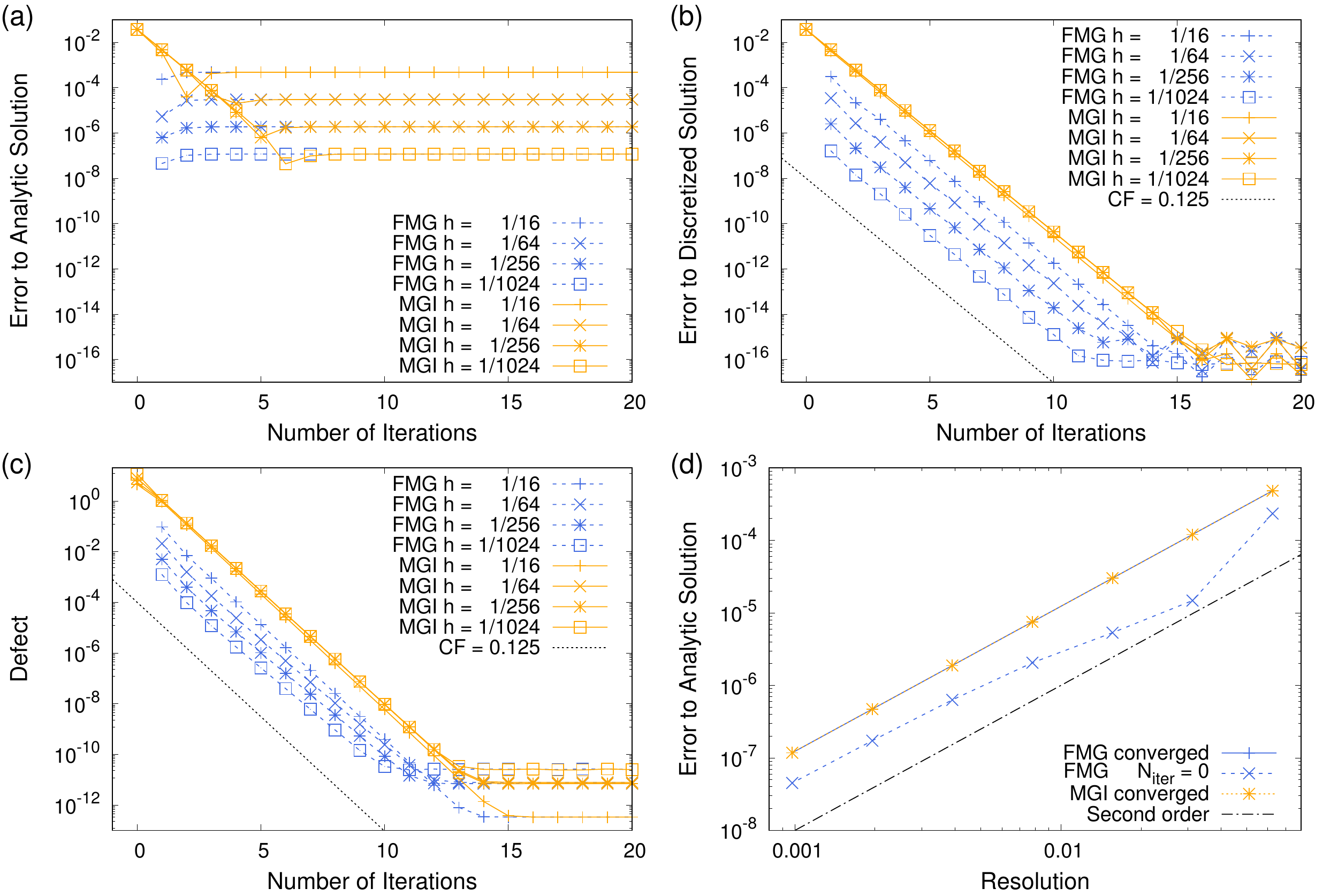}
\end{center}
\caption{The convergence behavior for the sinusoidal wave test on uniform grids. (a-c) The volume-weighted root-mean-square of the errors and defect versus the number of iterations. The blue lines and orange lines show the results of the FMG and MGI algorithms with various resolutions. \rev{The zeroth iteration points for MGI are corresponding to the initial guess, while the first iteration points for FMG are the results after the first FMG sweep.} The black dotted line shows the convergence factor of 0.125. (d) The error compared to the analytic solution versus the resolution. The black dash-dotted line shows second-order convergence.}\label{fig:uniformconvergence}
\end{figure*}

\section{Test Results}\label{sec:tests}
In order to demonstrate the accuracy and performance of our multigrid solver, we show results of some test problems. \rev{First we demonstrate accuracy and performance of the gravity solvers on uniform grids (Section \ref{sec:uniformtest}) and with AMR (Section \ref{sec:binary}). We also discuss the Jeans wave test in which hydrodynamics and self-gravity are coupled (Section \ref{sec:jeans}). Lastly, we show collapse of a magnetized molecular cloud in Section \ref{sec:collapse} as a practical example of self-gravitational MHD simulations with AMR.}. All the performance tests are carried out on Cray XC50 Supercomputer at Center of Computational Astrophysics, National Astronomical Observatory of Japan. Each node is equipped with two Intel Xeon Skylake 6148 processors (2.4GHz, 40 cores per node) and connected with the Aries interconnect. We compile the code using Intel C++ Compiler 2018 and Cray MPI version 7.7.0, and we report results with flat MPI parallelization.

\subsection{Uniform Grid: Sinusoidal Waves}\label{sec:uniformtest}
We first demonstrate the behavior of the code using a simple static problem. Here we solve the gravitational potential for a sinusoidal density distribution in a periodic box. The computational domain size is $L_x\times L_y\times L_z$ and all the boundary conditions are periodic. The density is set up overlaying sinusoidal waves in the three dimensions, 
\begin{eqnarray}
\rho(x,y,z) &=& \rho_0+A \sin\left(\frac{2\pi x}{L_x}\right)\sin\left(\frac{2\pi y}{L_y}\right)\sin\left(\frac{2\pi z}{L_z}\right),
\end{eqnarray}
where $\rho_0$ is the background density and $A$ is the amplitude. This problem has an analytic solution:
\begin{eqnarray}
\phi(x,y,z) = -\frac{4\pi G A}{(2\pi/L_x)^2+(2\pi/L_y)^2+(2\pi/L_z)^2}\nonumber\\
\times\sin\left(\frac{2\pi x}{L_x}\right)\sin\left(\frac{2\pi y}{L_y}\right)\sin\left(\frac{2\pi z}{L_z}\right) + \phi_0,
\end{eqnarray}
where we set the zero point of the potential to $\phi_0=0$ as discussed in Section~\ref{sec:boundary}. In this test, we set $A=1$ and $\rho_0=2$ and take the unit of $G=1$.

\subsubsection{Convergence}\label{sec:uniformconvergence}
In order to see the convergence behavior of the FMG and MGI algorithms, we perform this test with a fixed box size of $L_x=L_y=L_z = 1$ changing the resolution from $h=1/16$ to $h=1/1024$. For MGI, we use $\phi(x,y,z) =0$ as the initial guess, which is na\"ive but not very poor in this case. We measure three quantities as functions of \rev{the number of the iterations}: the error compared to the pointwise analytic solution $\epsilon$ (equation~\ref{eq:trueerror}), the error compared to the fully-converged discretized solution $\delta$ (equation~\ref{eq:error}), and the defect $d$ (equation~\ref{eq:defect}). It should be noted that the errors and the defect have different dimensions. We use the solution after 20 iterations as the fully-converged discretized solution. The results are shown in Figure~\ref{fig:uniformconvergence}.

As shown in Panel (a), all the algorithms quickly reach saturation against the analytic solution. This is because the numerical solutions converge not to the analytic solution corresponding to infinite resolution but to the exact discretized solution for the finite resolution, as shown in Panel (b) \citep[see also][]{moon2019}. In all the algorithms, each V-cycle iteration reduces the error (b) and defect (c) by the convergence factor $CF \sim 0.125$, \rev{which is consistent with that obtained in literature \citep[e.g.][]{yavneh1996}}. Depending on the resolution and algorithms, 10--15 iterations are needed to reach fully converged solutions.

When the numerical solutions reach convergence, both FMG and MGI achieve second-order accuracy as shown in Panel \rev{(d)}. \rev{$N_{\rm iter}$ is the number of the additional V-cycle iterations after the first FMG sweep.} It is remarkable that the first FMG sweep \rev{without additional V-cycle iterations} can achieve the error as small as the truncation error (in this particular problem, the error after the first FMG sweep is smaller than the error of the converged solution). However, because the solution after the first FMG sweep still contains high-frequency noises originating from the RBGS iteration pattern, it is recommended to apply a few V-cycle iterations which can damp the high-frequency noises quickly. On the other hand, it is not necessary to repeat iterations until the defect reaches saturation. In practice, a few additional V-cycles should be enough to achieve sufficiently small potential error, e.g. $\delta\sim 10^{-6}$.

The cost of one FMG sweep is comparable to two V-cycle iterations because the operations on the finest level dominate the cost (Figure~\ref{fig:cycles}). Unless the initial guess is very close to the true solution, MGI costs more than FMG to reach the same error level. As shown in Figure~\ref{fig:uniformconvergence}, FMG is more effective in higher resolutions. In practice, we can use the potential from the last timestep as the initial guess in hydrodynamic simulations. However, we find that it is usually insufficient and FMG almost always outperforms MGI.

\subsubsection{Performance}\label{sec:uniformperformance}
To demonstrate the parallel performance of our implementations, we perform weak-scaling tests using the same problem. We show the results of FMG with $N_{\rm iter}=0$ and $N_{\rm iter}=10$, and compare the results with the performance of Athena++'s MHD solver and gravity solver using FFT. Here we fix the size of each MeshBlock and increase the domain size as we increase the number of parallel processes. As we measure the scaling using whole nodes with 40 cores, the computational domain is not always a cube but generally a cuboid, which is not necessarily an optimal configuration for the multigrid solvers because the coarsest level cannot be always reduced to a single cell. It is also not optimal for FFT because it works best if the number of cells is power of two. Nevertheless, this configuration is close to practical use cases.

\begin{figure}[t]
\begin{center}
\includegraphics[width=\columnwidth]{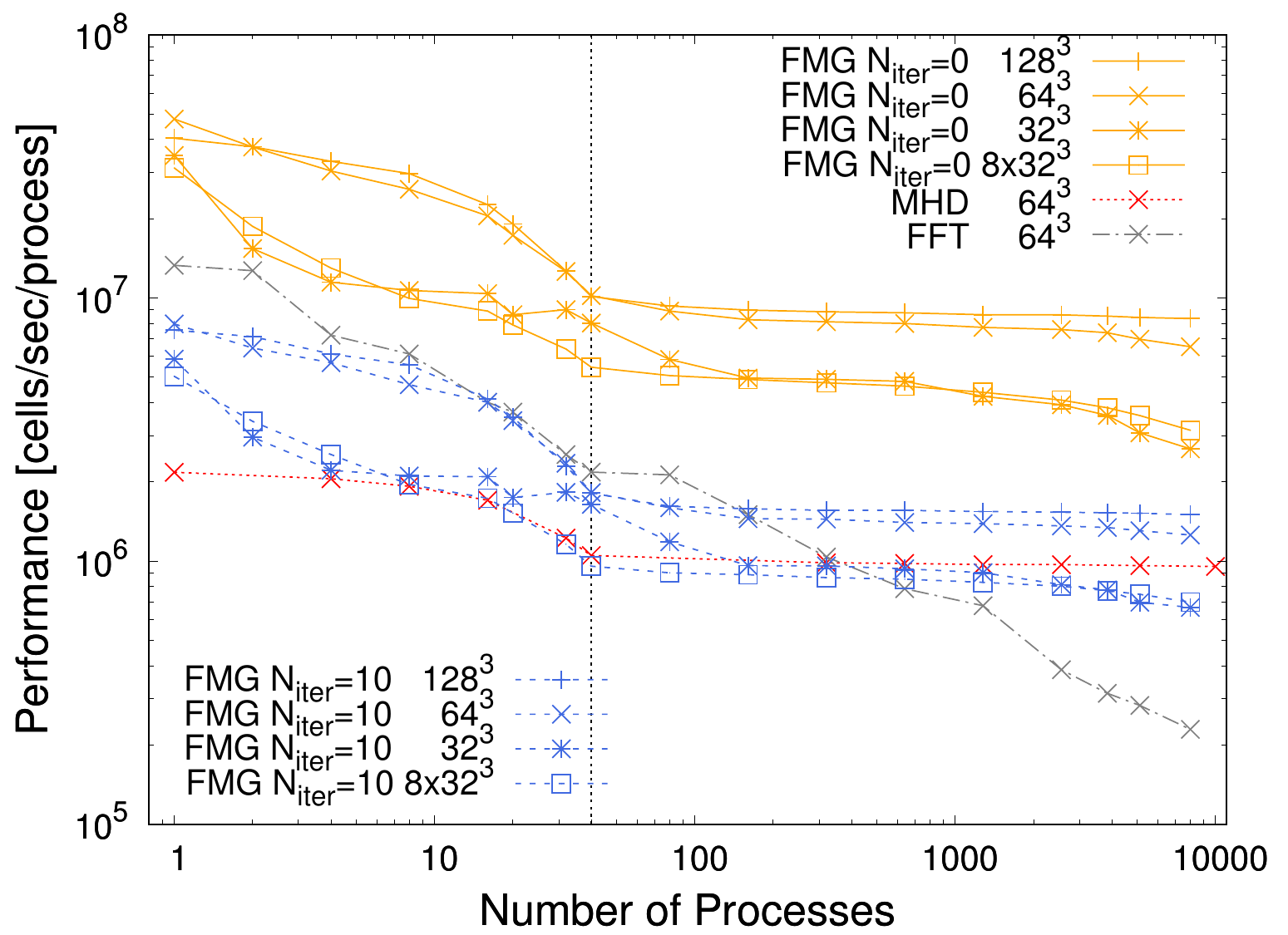}
\end{center}
\caption{Weak scaling test of the multigrid solvers on uniform grids with different \texttt{MeshBlock} sizes \rev{(higher is better)}. Orange solid lines: one sweep of FMG. Blue dashed lines: FMG with 10 V-cycle iterations. Red dotted line: MHD with \texttt{MeshBlocks} of $64^3$. Gray dash-dotted line: FFT with \texttt{MeshBlocks} of $64^3$. The vertical dotted line corresponds to one full node containing 40 cores.  Note FMG significantly outperforms FFTs, and is only a small fraction of the compute time for MHD.}\label{fig:uniformperformance} 
\end{figure}

The results of the weak scaling test are shown in Figure~\ref{fig:uniformperformance}. Because the performance of the multigrid solver is mainly limited by the memory bandwidth, it is inevitable that the performance declines as the number of processes increases within one node. Beyond one node, the runs with larger MeshBlocks scale better, and scalabilities between 2 nodes and 200 nodes are 89.6\%, 73.2\%, 45.7\% with MeshBlocks of $128^3$, $64^3$, $32^3$ cells for FMG one sweep and 92.9\%, 78.8\%, 56.3\% for FMG with 10 V-cycles. We also run a test with 8 MeshBlocks of $32^3$ cells, where the total computation size is the same as the $64^3$ case while the performance is similar to the $32^3$ case. We find that the performance of the multigrid solver fluctuates considerably depending on conditions of the network such as the distance between allocated nodes on the network and loads by other applications sharing the network. This is because the solver transfers many small messages and therefore it is sensitive to the network latency and bandwidth. The numbers reported here are the averages of the ten best results \rev{out of 100 runs}. \rev{We tend to get poor performance when the supercomputer is crowded, and it is up to $\sim 30\%$ slower in the worst cases.} FMG with 10 V-cycles costs about 5--6 times more than one sweep of FMG because one sweep of FMG costs about two V-cycle iteration as discussed in the previous section.

It is interesting to compare the performance of the multigrid solver to the FFT solver. As shown in the Figure, the FFT solver is not very scalable and the performance degrades with larger simulation sizes. This is not because there is a problem in the implementation of the FFT solver in Athena++, but it is actually a natural consequence because FFT's cost is intrinsically $O(N\log N)$. Therefore, the multigrid solver outperforms the FFT solver in practical use cases, particularly for massively parallel simulations.

Compared to the MHD solver using the same MeshBlocks of $64^3$, one FMG sweep costs only about 15\% while FMG with 10 V-cycles costs about 70\% of the MHD solver. Considering that we need to call the gravity solver twice per timestep with the second-order time integrator, these results indicate that the cost of the gravity solver on uniform grids is considerably smaller if only one FMG sweep is used, and it is comparable to MHD even with 10 V-cycles. In practical use cases, we need to apply the V-cycle iterations only a few times, therefore the self-gravity should cost only a fraction of the MHD solver. The parallel scalability of our gravity solver is not perfect because the self-gravity intrinsically requires extensive global communications, but these results demonstrate that our solver is highly efficient and enables large-scale self-gravitational hydrodynamic simulations.

\begin{figure*}[t]
\begin{center}
\includegraphics[width=\textwidth]{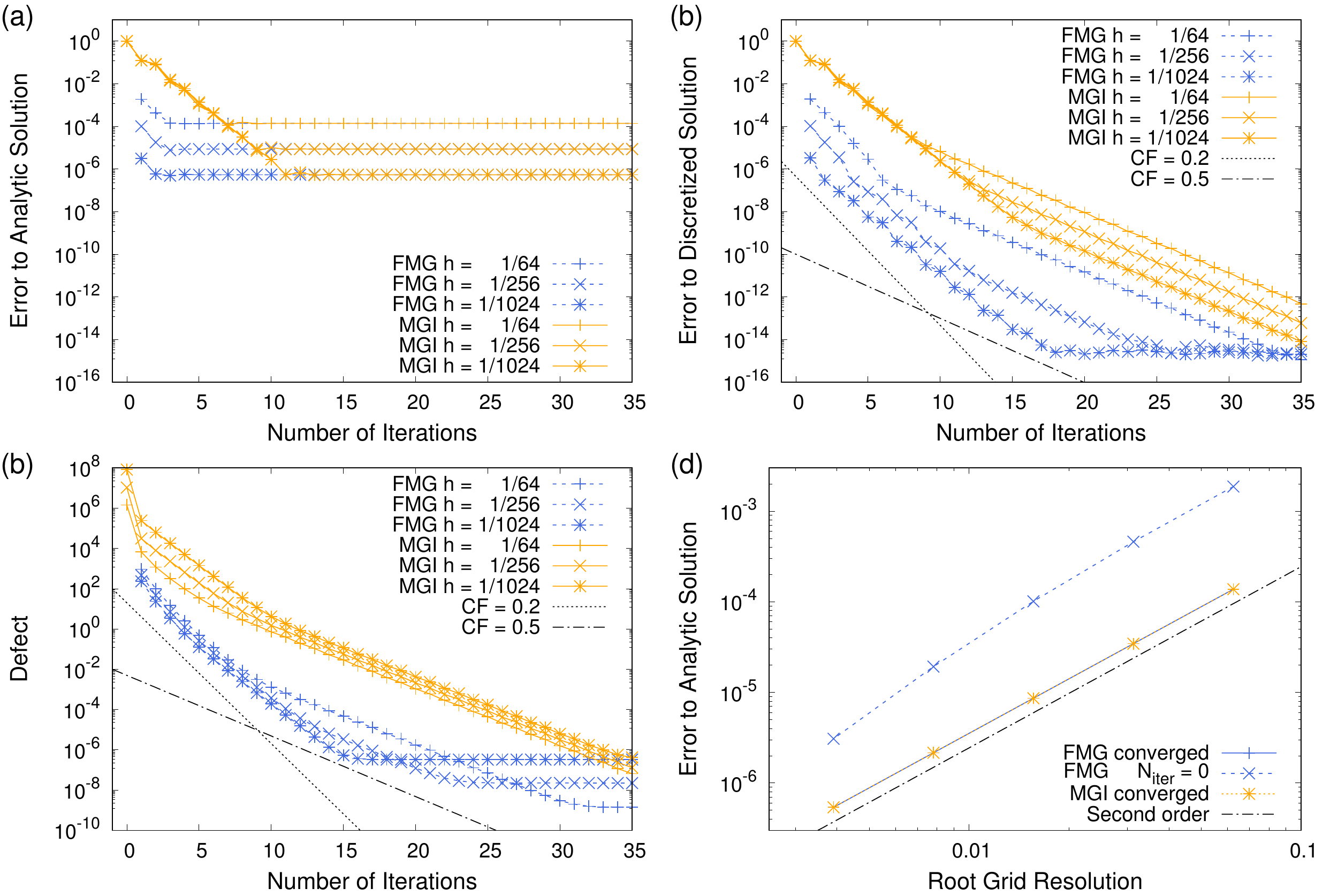}
\end{center}
\caption{Same as Figure~\ref{fig:uniformconvergence} but for the convergence behavior for the binary potential test with AMR. \rev{The errors in Panels (a) and (b) are normalized as in equations~(\ref{eq:normtrueerr}) and (\ref{eq:normdiscerr}).}}
\label{fig:amrconvergence}
\end{figure*}
\subsection{Mesh Refinement: Binary Potential}\label{sec:binary}
Next, we demonstrate the accuracy and performance of our multigrid solvers with AMR. As a non-trivial but simple static problem, we solve the potential of an unequal-mass binary system. We put the primary star with the mass of $M_1 = 2$ at $(x_1,y_1,z_1) = (6/1024, 0, 0)$ and the secondary with $M_2 = 1$ at $(x_2,y_2,z_2) = (-12/1024, 0, 0)$. Each star is modeled as a uniform sphere with the radius of $R=6/1024$. In order to make the density distribution near the surface smooth, we split each cell into $10^3$ subcells and calculate the cell density by taking the average over the subcells. The isolated boundary conditions using the multipole expansion up to hexadecapoles are applied on all the boundaries. The analytic solution of this problem is given as follows:
\begin{eqnarray*}
\phi(x,y,z) &=& \phi_1+ \phi_2,\\
\phi_1 &=& \left\{
\begin{aligned}
-\frac{GM_1}{\Delta r_1}\hspace{7.5em}& (\Delta r_1 \geq R)\\
-\frac{GM_1}{2R^3}\left[3R^2-(\Delta r_1)^2\right]\hspace{2em}& (\Delta r_1 < R)
\end{aligned}\right.\\
\phi_2 &=& \left\{
\begin{aligned}
-\frac{GM_2}{\Delta r_2}\hspace{7.5em}& (\Delta r_2 \geq R)\\
-\frac{GM_2}{2R^3}\left[3R^2-(\Delta r_2)^2\right]\hspace{2em}& (\Delta r_2 < R)
\end{aligned}\right.\\
\Delta r_1 &=& \sqrt{(x-x_1)^2 + (y-y_1)^2 + (z-z_1)^2},\\
\Delta r_2 &=& \sqrt{(x-x_2)^2 + (y-y_2)^2 + (z-z_2)^2}.
\end{eqnarray*}

The size of the computing domain is $[-0.5,0.5]\times [-0.5,0.5] \times [-0.5,0.5]$, and we place  refined grids with $2^l$ times higher resolution at $[-0.5^l,0.5^l]\times [-0.5^l,0.5^l] \times [-0.5^l,0.5^l]$ up to $l=4$ self-similarly. 

\begin{figure*}[t]
\begin{center}
\includegraphics[width=0.9\textwidth]{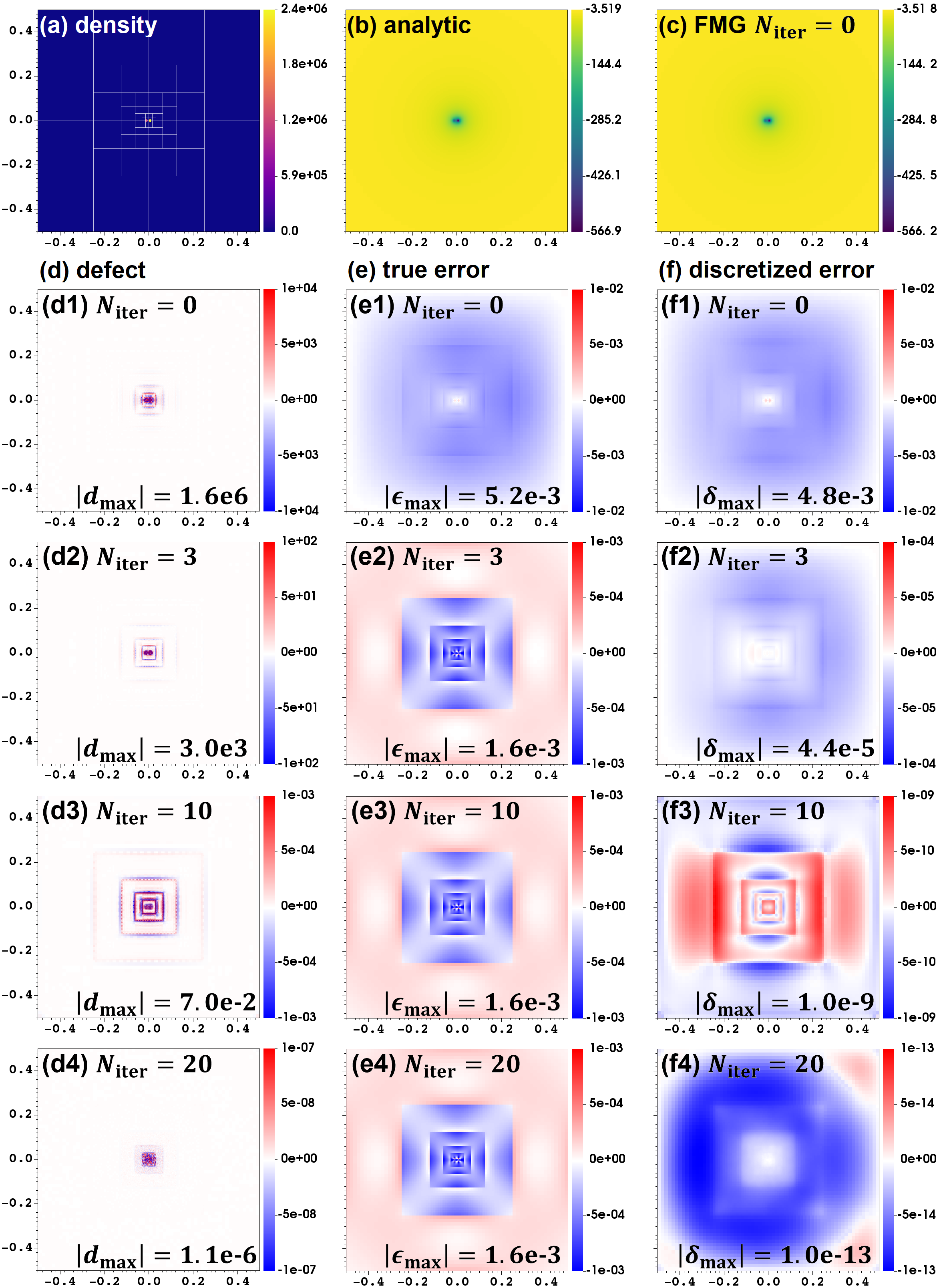}
\end{center}
\caption{The results of the binary potential test with the root grid resolution of $h=1/64$. (a) The density and \texttt{MeshBlock} distribution. (b) The analytic solution. (c) The result obtained with just one sweep of FMG. (d) The defect. (e) The error compared to the analytic solution. (f) The error compared to the fully-converged discretized solution. The ranges of the color bars are adjusted to clearly show the distributions, and the maximum values of the errors and defect are shown in the panels.}
\label{fig:binary}
\end{figure*}
\subsubsection{Convergence}\label{sec:amrconvergence}
We perform a similar resolution study as in Section~\ref{sec:uniformconvergence}. We change the resolution of the root grid from $h=1/64$ to $h=1/1024$. The maximum resolution on the finest level is 16 times higher than that of the root grid. For lower resolutions, the binary system is completely unresolved on the root grid. Again we use $\phi(x,y,z) =0$ as the initial guess for MGI, which is very poor in this case. We use the solution after 40 iterations as the fully-converged discretized solution. The results are shown in Figure~\ref{fig:amrconvergence}. In this test, because the potential in this problem drastically varies in space, we measure the potential error normalized by the analytic solution instead of the potential itself as in the previous test:
\begin{eqnarray}
\bm{\delta}_{\rm normalized}&=&\frac{\bm{\phi}^*_h-\bm{\phi}^n}{\bm{\phi}^*_h},\label{eq:normdiscerr}\\
\bm{\epsilon}_{\rm normalized}&=&\frac{\bm{\phi}^*-\bm{\phi}^n}{\bm{\phi}^*},\label{eq:normtrueerr}
\end{eqnarray}
where $\bm{\phi}^n$ is the numerical solution after the $n$-th V-cycle iteration, $\bm{\phi}^*_h$ the fully-converged discretized solution and $\bm{\phi}^*$ the analytic solution, respectively. The root-mean-square of the errors plotted in Figure~\ref{fig:amrconvergence} can be considered as the typical fractional error. On the other hand, we plot the defect itself and it is not normalized.

Overall, the solvers behave similarly as in the uniform grid case, and they achieve second-order accuracy as shown in Panel (d). However, the detail of the convergence behavior is different. While FMG quickly reach saturation against the analytic solution after only a few iterations, MGI takes considerably more iterations as the initial guess is not good (Panel (a)). The error compared to the fully-converged discretized solution (b) and the defect (c) exhibit the detailed behaviors of the algorithms. Initially, FMG converges quickly with $CF \sim 0.2$, but later it slows down to $CF \sim 0.5$. MGI shows a similar trend, but its initial convergence is slower than FMG. This is because the first sweep of FMG gives a good initial guess globally compared to the na\"ive guess used in MGI. These results clearly demonstrate the superiority of FMG over MGI.

To understand the convergence behavior, we show the distributions of the errors and defect of FMG in Figure~\ref{fig:binary}. The checker-board pattern seen in the defect originates from the RBGS smoothing operation. The error compared to the analytic solution $\epsilon$ quickly converges and it is small enough even after the first FMG sweep (Panels e1--e4). During the initial fast converging phase, the error to the fully converged solution $\delta$ and defect $d$ are reduced globally (Panels d1--d2 and f1--f2). In the later phase, the persisting errors at the level boundaries cause the slower convergence (Panels d3 and f3), but eventually the errors and defect are sufficiently reduced and approach the fully-converged solution (Panels d3 and f3). As seen in Panels (b), (c) and (e), the potential error against the analytic solution declines quickly even with a few iterations. Therefore, again, a few additional V-cycles should be enough to achieve sufficiently small error in practice.

\subsubsection{Performance}\label{sec:amrperformance}
We measure the parallel performance of our multigrid solver using the same problem. We measure strong scaling where the computational load of the whole simulation is fixed and weak scaling where the computational load per process is fixed. 

In order to simulate typical use cases with AMR, we apply V-cycle iterations three times after the first FMG sweep. To see the performance with different \texttt{MeshBlock} sizes, we run the test calculations with $16^3$ and $32^3$ cells per \texttt{MeshBlock}. In this test, we use only 36 cores per node so that the number of the \texttt{Multigrids} can be evenly divisible. In the strong scaling test, we set up the root grid of $256^3$ with $16^3$ \texttt{MeshBlocks} and $512^3$ with $32^3$ \texttt{MeshBlocks}, and 4 AMR levels are added self-similarly. This produces 18,432 \texttt{MeshBlocks} in total. We measure the performance from 36 processes (512 \texttt{MeshBlocks} per process) to 2,304 processes (8 \texttt{MeshBlocks} per process). In the weak scaling test, we use the same grid configuration but reduces the resolution so that each process always owns 8 \texttt{MeshBlocks}. The results as well as comparison with the performance of the MHD solver using the same \rev{static grid} configuration are shown in Figure~\ref{fig:amrperformance}.

\begin{figure}[t]
\begin{center}
\includegraphics[width=\columnwidth]{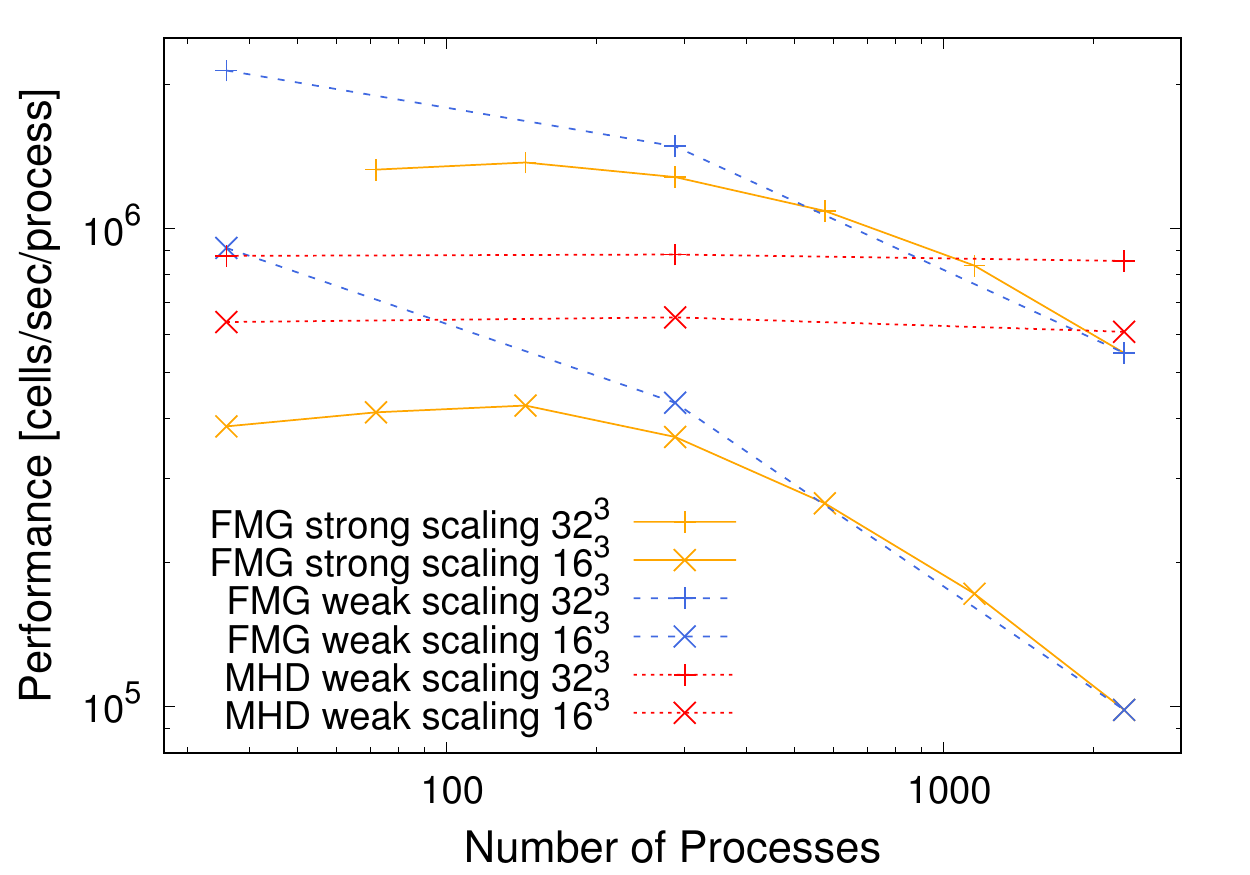}
\end{center}
\caption{The strong and weak scaling tests on AMR with FMG and $N_{\rm iter}=3$ \rev{(higher is better)}. Blue dashed lines: the strong scaling of the multigrid solver. Orange solid lines the weak scaling of the multigrid solver. Red dotted lines: the weak scaling of the MHD solver for comparison. The strong scaling runs with 36 cores and $32^3$ \texttt{MeshBlocks} are not shown because they exceed the memory limit per node.}
\label{fig:amrperformance}
\end{figure}

In this test, the parallel performance degrades with the large number of processes, particularly with small \texttt{MeshBlock} size. With 2,304 processes, the multigrid solver is about 2--6 times slower and less scalable compared to the MHD part for the same \texttt{MeshBlock} size. These ratios are not very good, but are actually comparable to the gravity solver of other AMR codes \citep[e.g.][]{flashgrav}. This is mainly because our solver on the \texttt{RootGrid} and \texttt{Octet} levels is not parallelized and their costs become significant. It is of crucial importance to choose appropriate grid configurations for good performance with AMR. \rev{In practice, it is reasonable to try \texttt{MeshBlocks} of $16^3$ cells first \citep[see also discussions in][]{athenapp}.}

\subsection{Dynamical Test: Jeans Wave}\label{sec:jeans}
As one of the simplest examples of hydrodynamic problems with self-gravity, here we consider the Jeans wave test is a periodic box with $L_x=1, L_y=0.5, L_z=0.5$. The initial density distribution is uniform with a sinusoidal perturbation:
\begin{eqnarray}
\rho(\bm{x}) &=& \rho_0+A \sin\left(\bm{k}\cdot\bm{x}\right),
\end{eqnarray}
where $\bm{k}$ is the wave number vector and $A=10^{-6}$ is the amplitude of the perturbation. We use $\bm{k}=\left(\frac{2\pi}{L_x}, \frac{2\pi}{L_y}, \frac{2\pi}{L_z}\right)$ so that the perturbation is periodic in all the directions. The corresponding wavelength of the perturbation is $\lambda=1/3$. We use the adiabatic equation-of-state with the adiabatic index $\Gamma=5/3$, and the initial sound speed is $c_{\rm s} = \sqrt{\frac{\Gamma  p_0}{\rho_0}}$. We set the initial density and pressure to $\rho_0 = 1$ and $p_0$=1, and the perturbation is isothermal. The dispersion relation of the system is 
\begin{eqnarray}
\omega^2=c_{\rm s}^2k^2-4\pi G\rho_0.
\end{eqnarray}
This perturbation is unstable when its wavelength exceeds the Jeans wavelength $\lambda_{\rm Jeans}\equiv\sqrt{\frac{\pi c_s^2}{G\rho_0}}$. We set the gravitational constant $G$ using the ratio between the wavelength of the perturbation and the Jeans wavelength $n_{\rm Jeans}\equiv\frac{\lambda}{\lambda_{\rm Jeans}}$ as an input parameter. For a stable initial condition ($n_{\rm Jeans} < 1$), the initial momentum is set to zero, while we give a corresponding momentum perturbation parallel to the wavenumber vector
\begin{eqnarray}
\rho\bm{v}(\bm{x})=\frac{\bm{k}}{k}\frac{|\omega|}{k}\rho_0A\cos\left(\bm{k}\cdot\bm{x}\right),
\end{eqnarray}
for an unstable case. For the hydrodynamic part, we use the second-order piecewise linear method for spatial reconstruction and the second-order van Leer time integrator with the Harten-Lax-van Leer Contact (HLLC) approximate Riemann solver.

\begin{figure}[t]
\begin{center}
\includegraphics[width=\columnwidth]{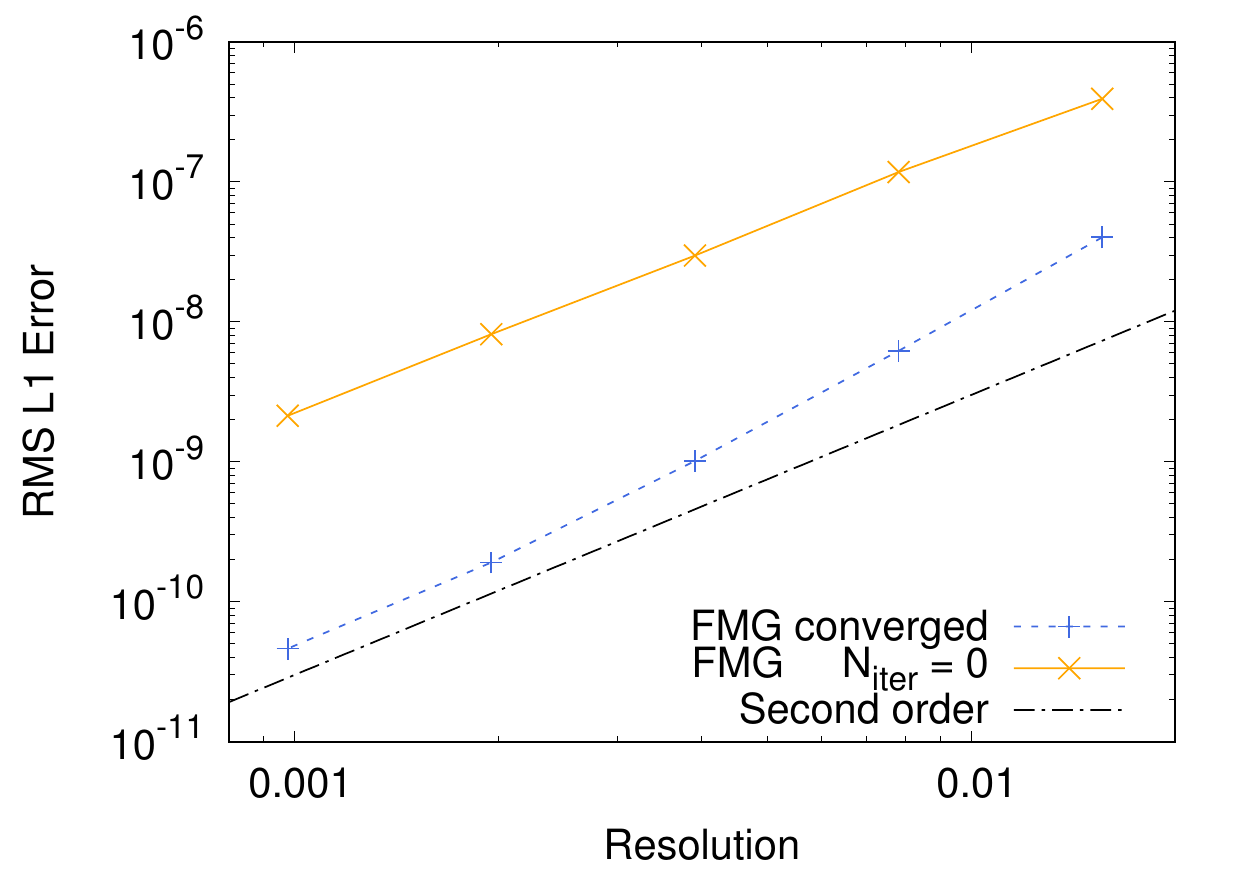}
\end{center}
\caption{The error compared to the analytic solution of the stable Jeans wave test. The blue dashed line shows the result using fully-converged solutions for self gravity, while the orange slid line shows the result using one sweep of FMG for self-gravity. The black dash-dotted line shows the second-order convergence.}
\label{fig:jeansconvergence}
\end{figure}

In Figure~\ref{fig:jeansconvergence}, we demonstrate the convergence of the stable case with $n_{\rm Jeans} = 0.5$ changing the resolution. \rev{Here we} compare fully-converged solutions \rev{and} just one sweep of FMG using the root-mean-square of the L1 errors of all the \rev{conservative} hydrodynamic variables, \rev{$(\rho, \rho v_x, \rho v_y, \rho v_z, E)$ where $E$ is the total energy density.}. Both methods achieve roughly second order accuracy. As expected, use of the fully-converged solutions improves the accuracy. While the error with $N_{\rm iter}=0$ is precisely second order, the error with the fully-converged solution slightly deviates from the second order convergence. This indicates that the error is dominated by the gravitational potential with $N_{\rm iter}=0$, but using the fully-converged solution it is dominated by the hydrodynamic part which is not fully second-order accurate because of the slope limiter. 

\begin{figure}[t]
\begin{center}
\includegraphics[width=\columnwidth]{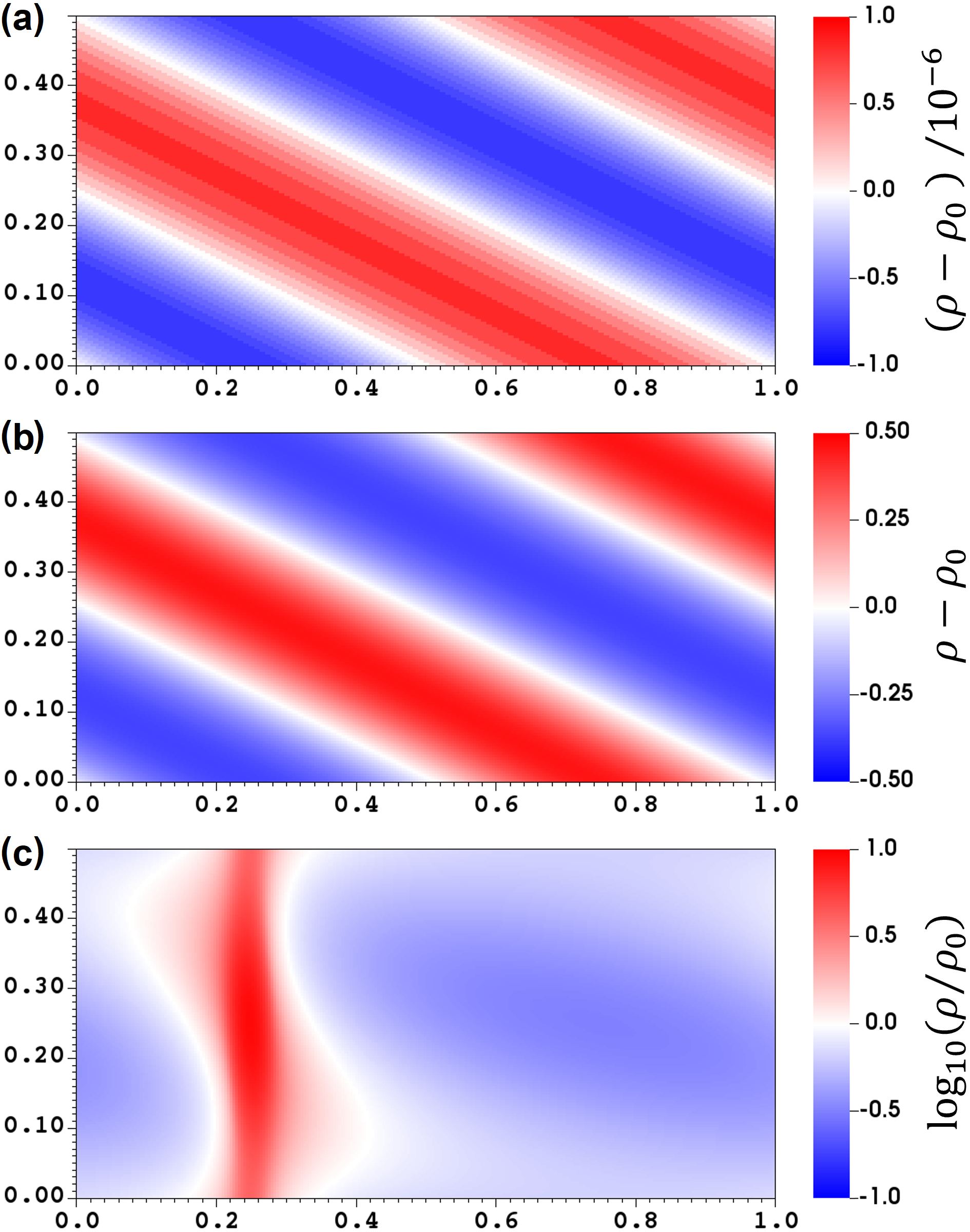}
\end{center}
\caption{The results of the Jeans wave test at $t=1.5$. (a) A stable case with $n_{\rm Jeans}=0.5$ using FMG with $N_{\rm iter}=0$. (b) An unstable case with $n_{\rm Jeans}=1.1$ using fully-converged gravitational potential. (c) An unstable case with $n_{\rm Jeans}=1.1$ using FMG with $N_{\rm iter}=0$. Note that the plotted quantities and ranges of the color bars are different, but in all the panels red color indicates density increase and blue does density decrease compared to the initial condition. The stripe pattern in Panel (a) is not due to the numerical error in the gravity solver but is due to loss of significance by subtraction between close values because the density perturbation is much smaller compared to the background density.}
\label{fig:jeans}
\end{figure}

Next, we compare the behaviors of the accuracy of the self-gravity solver using both stable and unstable solutions in Figure~\ref{fig:jeans}. In the stable case, FMG with $N_{\rm iter}=0$ and fully-converged solutions both reach qualitatively the same solutions (Panel (a)). However, in the unstable cases, they result in qualitatively different solutions. While the wave grows in the amplitude while maintaining the symmetry when fully-converged solutions are used (Panel (b)), it collapses into fragmentations with $N_{\rm iter}=0$ (Panel (c)). In the latter case, the gravitational instability grows faster and the symmetry breaks earlier due to the noises arising from the remaining errors in the gravitational potential. This test shows that it is important to control the error of the gravity solver when the system is sensitive to noises.

\subsection{Example Application: Collapse of a Magnetized Cloud}\label{sec:collapse}
Lastly, as a demonstration of the use of self-gravity in a MHD simulation with AMR, we present collapse of a magnetized cloud, which is a model of star formation in an isolated molecular cloud core. 

\begin{figure*}[t]
\begin{center}
\includegraphics[width=\textwidth]{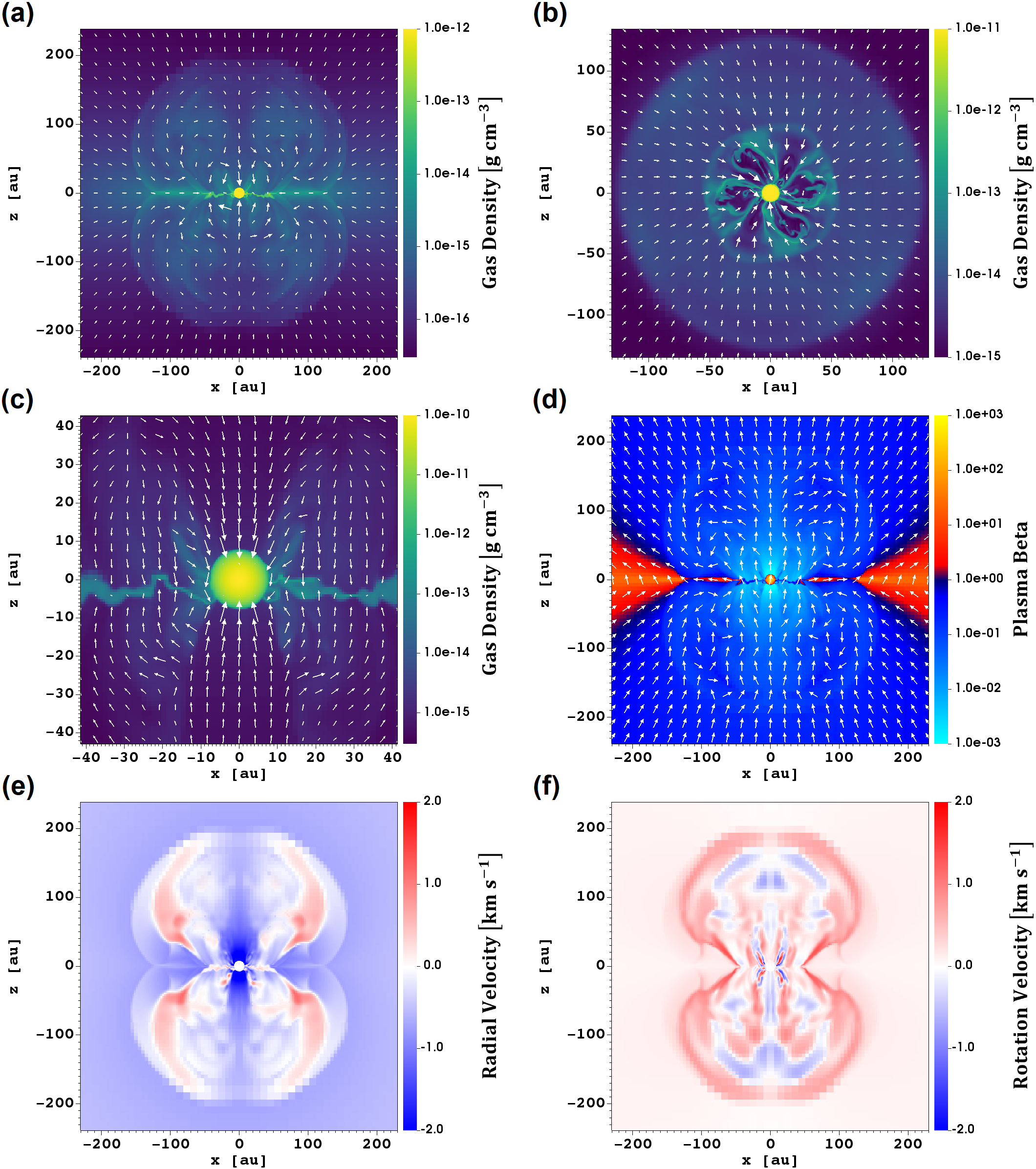}
\end{center}
\caption{Results of a MHD protostellar collapse simulation with $\rho_{\rm c}=10^{-8}\,{\rm g\,cm^{-3}}$. (a) Poloidal density slice with velocity vectors on large scales. (b) Equatorial density slice with velocity vectors on scales of the pseudo-disk. (c) Poloidal density slice with velocity vectors on scales of the first core. (d) Plasma beta with magnetic field direction vectors on large scales. (e) Radial velocity on large scales. (f) Rotational velocity on large scales. Note the length of velocity vectors in panels (a)-(c) correspond to different speeds.}
\label{fig:collapse}
\end{figure*}

As an initial condition, we adopt a critical Bonnor-Ebert sphere \citep{bonnor,ebert} with temperature $T=10\,{\rm K}$ and make it unstable by multiplying the density by an enhancement factor $f=1.2$. We also add 10\% $m=2$ sinusoidal perturbation in the density. The total mass of the cloud is set to $M=1M_\odot$, and the initial cloud radius and the density at the center of the cloud are $R=8,400\,{\rm au}$ and $\rho_0=1.13\times 10^{-18}\, {\rm g\, cm^{-3}}$, respectively. We introduce solid-body rotation and a uniform magnetic field both aligned to the $z$-axis. The angular velocity is $\omega=5.55\times 10^{-14}\, {\rm s^{-1}}$, corresponding to $\omega t_{\rm ff} = 0.1$ where $t_{\rm ff}=\frac{3\pi}{32G\rho_0}=5.71\times 10^4 \,{\rm yrs}$ is the initial free-fall time at the center of the cloud, and the magnetic field is $B_z = 27.4\,{\rm \mu G}$, corresponding to the normalized mass-to-flux ratio of $\mu\equiv \frac{M/\Phi}{(M/\Phi)_{\rm crit}}=3$ where $(M/\Phi)_{\rm crit}=\frac{0.53}{3\pi}\left(\frac{5}{G}\right)^{1/2}$ is the critical mass-to-flux ratio for a spherical cloud \citep{ms76}.

The simulation domain is $[-10,400\,{\rm au}:10,400\,{\rm au}]^3$, and the root grid is resolved with $128^3$ cells. We use AMR with \texttt{MeshBlocks} consisting of $16^3$ cells to resolve the local Jeans length at least with 32 cells. The boundary conditions are set to model a cloud confined by the ambient gas pressure. The velocity outside the initial cloud radius is fixed to zero. For self-gravity, isolated boundary conditions computed from a multipole expansion up to hexadecapoles are used, ignoring the mass outside the initial cloud radius to suppress the higher-order multipole components from the gas near the edges of the computing domain. 

For MHD, we use the second-order van-Leer time integrator and second-order Piecewise Linear Method for spatial reconstruction with the HLLD approximate Riemann solver. We adopt FMG with $N_{\rm iter}=10$ for the self-gravity.  To model the thermal evolution in a collapsing cloud, we adopt the barotropic relation:
\begin{eqnarray}
P=\rho c_{\rm s}^2\left[1+\left(\frac{\rho}{\rho_{\rm crit}}\right)^{2(\Gamma-1)}\right]^{1/2},
\end{eqnarray}
where $c_{\rm s}=0.19\, {\rm km\, s^{-1}}$ is the sound speed at $T=10\,{\rm K}$, $\rho_{\rm crit}=10^{-13}\,{\rm g\,cm^{-3}}$ the critical density, and $\Gamma=5/3$ the adiabatic index, respectively. With this formula, the gas is isothermal in the low-density region $\rho \ltsim \rho_{\rm crit}$ due to efficient radiation cooling, and it becomes adiabatic in the high-density region where radiation cooling is inefficient. These configurations are similar to those used in many previous MHD simulations of protostellar collapse \citep[e.g.][]{machida2004,hf2008,tomida2015,tomida2017} as well as in the method papers of the RAMSES \citep{ramsesmhd} and AREPO \citep{arepomhd} codes.

Figure~\ref{fig:collapse} displays the results at $t=1.96\times 10^5\, {\rm yrs}$ when the central density and temperature reach $\rho_{\rm c} = 10^{-8}\,{\rm g\,cm^{-3}}$ and $T_{\rm c}=1,000\,{\rm K}$. At this point, 9 levels of refinement are generated, and the corresponding finest resolution is $0.326\,{\rm au}$. A so-called first hydrostatic core (``first core" for short) is formed at the center of the cloud as the gas becomes adiabatic and the gas pressure and gravity balance. The first core is almost spherically symmetric as a result of highly efficient angular momentum transport by the magnetic field. The flattened disk-like density distribution around the first core is a so-called pseudo-disk, which is not supported by rotation but is radially infalling. Slow bipolar outflows with a wide opening angle and $v\sim 1\, {\rm km\, s^{-1}}$ are launched from the pseudo-disk by the magneto-centrifugal mechanism \citep{bp82}. The pseudo-disk is highly disturbed in the vicinity of the first core due to the magnetic interchange instability \citep{spruit1995,lam2019} because the magnetic field is strongly concentrated as a result of the gravitational collapse. These results are consistent with the results of the previous ideal MHD simulations \citep[e.g.][]{tomida2015}. This calculation required about 4,900 core-hours, or 30.7 hours using 160 cores on the Cray XC50 supercomputer. \rev{The gravity solver takes about 85--90\% of the computing time, which varies with the AMR grid structure.} It demonstrates that our code is highly capable and is applicable to practical astrophysical problems.

\section{Summary and Discussion}
In this paper, we have described our implementation of self-gravity solvers based on geometric multigrid algorithms. The solvers are built on top of the Athena++ framework and parallelized using its dynamic execution model using a \texttt{TaskList}. We have implemented both the multigrid iteration (MGI) as well as the full multigrid (FMG) scheme with V(1,1) cycles, and have used a variety of test problems to demonstrate the accuracy and performance of each. In general, we find that FMG is superior to MGI as it converges faster and does not need any initial guess.

Our test problems show that the multigrid solvers can quickly reach saturation to an analytic solution and achieve second order accuracy. In particular, just one sweep of FMG with no additional V-cycle iterations gives a remarkably good solution compared to the analytic solution. However, the defect and error compared to a fully-converged discretized solution are still large, including small-scale noise arising from the RBGS iteration pattern as well as AMR level boundaries. In practice it is not necessary to iterate until the solution is fully converged, however some additional V-cycle iterations are recommended to suppress the error and noise. We find that FMG with 3-10 additional V-cycle iterations should be sufficient in most cases, but this is problem dependent and therefore it is important to monitor and control error carefully.

While the performance of our multigrid solver is very good on uniform grids, with AMR the scaling depends strongly on the grid decomposition. In particular, the parallel performance quickly degrades when many small \texttt{MeshBlocks} are used because the cost of the non-parallelized portion of the algorithm becomes non-negligible. Therefore, it is important to configure the AMR grid to balance flexibility and performance. In order to improve parallel performance with AMR, an additional layer of parallelization for the \texttt{RootGrid} and \texttt{Octet} levels may be needed. Alternatively, heterogeneous parallelization in which a small fraction of the compute processes are assigned to multigrid while the rest update the hydrodynamics may improve the overall performance and scalabilty. Such enhancements will be explored in the future.

\rev{There are many implementations of self-gravity solvers on AMR in literature. Some of them adopt the level-by-level approach \citep[e.g.][]{art,ramsesmg,enzo,dispatchgrav}. One of the motivations of this approach is use of the individual timestepping with AMR. While it is straightforward to combine the level-by-level approach and individual timestepping, we should note that the gravity propagates instantaneously and we need globally-consistent solutions, even with proper temporal interpolation. This is why we adopt the combination of the uniform timestepping and our globally-consistent multigrid solver. Another difference among these implementations is the design of AMR; \citet{art}(ART) and \citet{ramsesmg}(RAMSES) are cell-based AMR, \citet{enzo} is patch-based, and \citet{nirvana}(NIRVANA), \citet{sfumato}(SFUMATO), \citet{flashmg}(FLASH)\footnote{Although it is referred as ``multigrid" in \citet{flashmg}, it is different from what is presented in this paper. In this scheme, they solve the Poisson equation in each block using a direct solver such as FFT, and apply corrections between blocks on different levels repeatedly until a globally-consistent solution is obtained.}, \citet{tk2019}(AMRVAC) as well as Athena++ adopt octree-block based AMR. It is straightforward to implement the multigrid algorithms on the octree-block based AMR because relations between blocks are limited, while it is more complicated to combine with the other AMR designs. Among them, our implementation is similar to that in the SFUMATO code \citep{sfumato}, which obtains globally-consistent solutions using the full multigrid method on the octree-block based AMR with uniform timestepping. A notable difference is that \citet{sfumato} applies flux-correction at level boundaries in order to satisfy the Gauss's theorem, while we use the mass-conservation forumula \citep{masscons} in the discretization which does not require any additional computation nor communication. In addition, our scheme is efficiently parallelized using the task-based dynamic execution model of Athena++.}

Multigrid algorithms are applicable to many physical processes in astrophysics and not limited to self-gravity. For example, radiation or cosmic-rays transport with moment-based method such as Flux-Limited Diffusion \citep{fld,zeusfld} or the Variable Eddington Tensor method \citep{zeusvet,athrt} using an implicit time integrator is of particular interest, and we are planning to develop such extensions. The Poisson equation for the pressure in incompressible hydrodynamics is another possible application. \rev{Similarly, multigrid can be applied to the Hodge-Helmholtz decomposition to extract compressive and solenoidal components from a vector field} \citep[e.g.][]{miniati2014,valles-perez2021}. We are also planning to implement the particle-mesh gravity solver with AMR combining the multigrid gravity solver and the particle module of Athena++. While our current implementation only supports Cartesian coordinates, extension to cylindrical, spherical polar and other coordinate systems is possible.  Finally, we have recently developed a performance-portable version of the Athena++ AMR framework that runs on a variety of heterogeneous computing architectures including GPUs (Stone et al., in preparation), and we plan to implement the multigrid solvers described here into this new version of the code as well.

The code is publicly distributed through the Athena++ GitHub repository \footnote{\url{https://github.com/PrincetonUniversity/athena}} under the BSD open-source license.

\begin{acknowledgements}
We thank Eve Ostriker, Chris White, Patrick Mullen, Chang-Goo Kim, Sanghyuk Moon, Kazunari Iwasaki, and Tomoaki Matsumoto for fruitful discussions. KT is grateful for the hospitality of Institute for Advanced Study and Department of Astrophysical Sciences at Princeton University, where most of this paper was written. This work is supported by Japan Society for the Promotion of Science (JSPS) KAKENHI Grant Numbers JP16H05998, JP16K13786, JP17KK0091, JP18H05440, JP21H04487 and JP22KK0043. Numerical computations were in part carried out on Cray XC50 at Center for Computational Astrophysics, National Astronomical Observatory of Japan. This work was supported by Ministry of Education, Culture, Sports, Science and Technology (MEXT) as ``Program for Promoting Researches on the Supercomputer Fugaku" (Toward a unified view of the universe: from large scale structures to planets, JPMXP1020200109) and used computational resources of supercomputer Fugaku provided by the RIKEN Center for Computational Science (Project IDs: hp200124, hp210164, hp220173). JS acknowledges support from NASA grant 80NSSC21K0496, and from the Eric and Wendy Schmidt Fund for Strategic Innovation.  We used VisIt \citep{visit} to produce Figures~\ref{fig:binary}, \ref{fig:jeans} and \ref{fig:collapse}.
\end{acknowledgements}
\software{Athena++ \citep{athenazenodo}}

\bibliography{ms}

\appendix
\section{Multipole Expansion Boundary Condition}\label{sec:multipole}
Here, we describe how to calculate isolated boundary conditions using the multipole expansion. The gravitational potential far away from the center of mass can be well approximated with the multipole expansion truncated at a finite order of the expansion, as the potential produced by higher order multipoles decline sharply at a large distance from the center of mass. In Athena++, we implement the multipole expansion up to hexadecapoles.

The gravitational potential at a point $\bm{r}=(x,y,z)$ outside the mass distribution is expanded as:
\begin{eqnarray}
\phi(\bm{r})=G\sum_{l=0}^{l_{\rm max}}\sum_{m=-l}^l\left(\frac{Q_{l,m}}{r^{l+1}}\right)\sqrt{\frac{4\pi}{2l+1}}Y_{l,m}(\bm{r}),
\end{eqnarray}
where $l_{\rm max}$ is the maximum order of the expansion, $r=\sqrt{x^2+y^2+z^2}$, and $Y_{l,m}$ is the real spherical harmonics. The multipole moments $Q_{l,m}$ are defined as 
\begin{eqnarray}
Q_{l,m}&=&\sum_V \rho(\bm{r}') \Delta V \sqrt{\frac{4\pi}{2l+1}}(r')^l\,Y_{l,m}(\bm{r}'),
\end{eqnarray}
where the summation is taken over the whole computing domain and $\Delta V$ is the volume of the cell at position $\bm{r}'$.

We list the real spherical harmonics in Cartesian upto $l=4$:
\begin{eqnarray*}
&Y_{0,0}=\sqrt{\frac{1}{4\pi}},&\\
&Y_{1,-1}=\sqrt{\frac{3}{4\pi}}\frac{y}{r},\hspace{1em}
Y_{1,0}=\sqrt{\frac{3}{4\pi}}\frac{z}{r},\hspace{1em}
Y_{1,1}=\sqrt{\frac{3}{4\pi}}\frac{x}{r},&\\
&Y_{2,-2}=\sqrt{\frac{15}{4\pi}}\frac{xy}{r^2},\hspace{1em}
Y_{2,-1}=\sqrt{\frac{15}{4\pi}}\frac{yz}{r^2},\hspace{1em}
Y_{2,0}=\sqrt{\frac{5}{16\pi}}\frac{3z^2-r^2}{r^2},\hspace{1em}
Y_{2,1}=\sqrt{\frac{15}{4\pi}}\frac{xz}{r^2},\hspace{1em}
Y_{2,2}=\sqrt{\frac{15}{4\pi}}\frac{x^2-y^2}{2r^2},&\\
&Y_{3,-3}=\sqrt{\frac{35}{32\pi}}\frac{y(3x^2-y^2)}{r^3},\hspace{1em}
Y_{3,-2}=\sqrt{\frac{105}{4\pi}}\frac{xyz}{r^3},\hspace{1em}
Y_{3,-1}=\sqrt{\frac{21}{32\pi}}\frac{y(5z^2-r^2)}{r^3},&\\
&Y_{3,0}=\sqrt{\frac{7}{16\pi}}\frac{z(5z^2-3r^2)}{r^3},&\\
&Y_{3,1}=\sqrt{\frac{21}{32\pi}}\frac{x(5z^2-r^2)}{r^3},\hspace{1em}
Y_{3,2}=\sqrt{\frac{105}{4\pi}}\frac{z(x^2-y^2)}{2r^3},\hspace{1em}
Y_{3,3}=\sqrt{\frac{35}{32\pi}}\frac{x(x^2-3y^2)}{r^3},&\\
&Y_{4,-4}=\sqrt{\frac{315}{4\pi}}\frac{xy(x^2-y^2)}{2r^4},\hspace{1em}
Y_{4,-3}=\sqrt{\frac{315}{32\pi}}\frac{yz(3x^2-y^2)}{r^4},\hspace{1em}
Y_{4,-2}=\sqrt{\frac{45}{16\pi}}\frac{xy(7z^2-r^2)}{r^4},&\\
&Y_{4,-1}=\sqrt{\frac{45}{32\pi}}\frac{yz(7z^2-3r^2)}{r^4},\hspace{1em}
Y_{4,0}=\sqrt{\frac{9}{256\pi}}\frac{35z^4-30z^2r^2+3r^4}{r^4},\hspace{1em}
Y_{4,1}=\sqrt{\frac{45}{32\pi}}\frac{xz(7z^2-3r^2)}{r^4},\\
&Y_{4,2}=\sqrt{\frac{45}{16\pi}}\frac{(x^2-y^2)(7z^2-r^2)}{2r^4},\hspace{1em}
Y_{4,3}=\sqrt{\frac{315}{32\pi}}\frac{xz(x^2-3y^2)}{r^4},\hspace{1em}
Y_{4,4}=\sqrt{\frac{315}{4\pi}}\frac{x^2(x^2-3y^2)-y^2(3x^2-y^2)}{8r^4}.&
\end{eqnarray*}

For the Poisson equation for self-gravity, all the dipole moments $Q_{1,m}$ are zero if the origin of the multipole expansion is the center of mass. In our experience, the best practice is usually to use the center of mass as the origin of the multipole expansion, because the higher order moments are also small when the center of mass is used as the origin. By default, Athena++ automatically calculates the center of mass at every timestep. Alternatively, we provide an option for users to specify the origin, which is useful when the center of mass does not move.

Note that the accuracy of the multipole expansion is limited by the maximum order of the expansion. The multipole expansion is reasonably accurate if the mass distribution is centrally condensed and far away from boundaries of the computational domain. However, if the mass distribution extends close to boundaries, it is less accurate. We plan to implement a more general method for isolated boundary conditions such as the James algorithm \citep{moon2019}.

\section{Comparison between Our Implementation and Level-by-Level Approach}\label{sec:amr_lbl}
\begin{figure*}[t]
\begin{center}
\includegraphics[width=\textwidth]{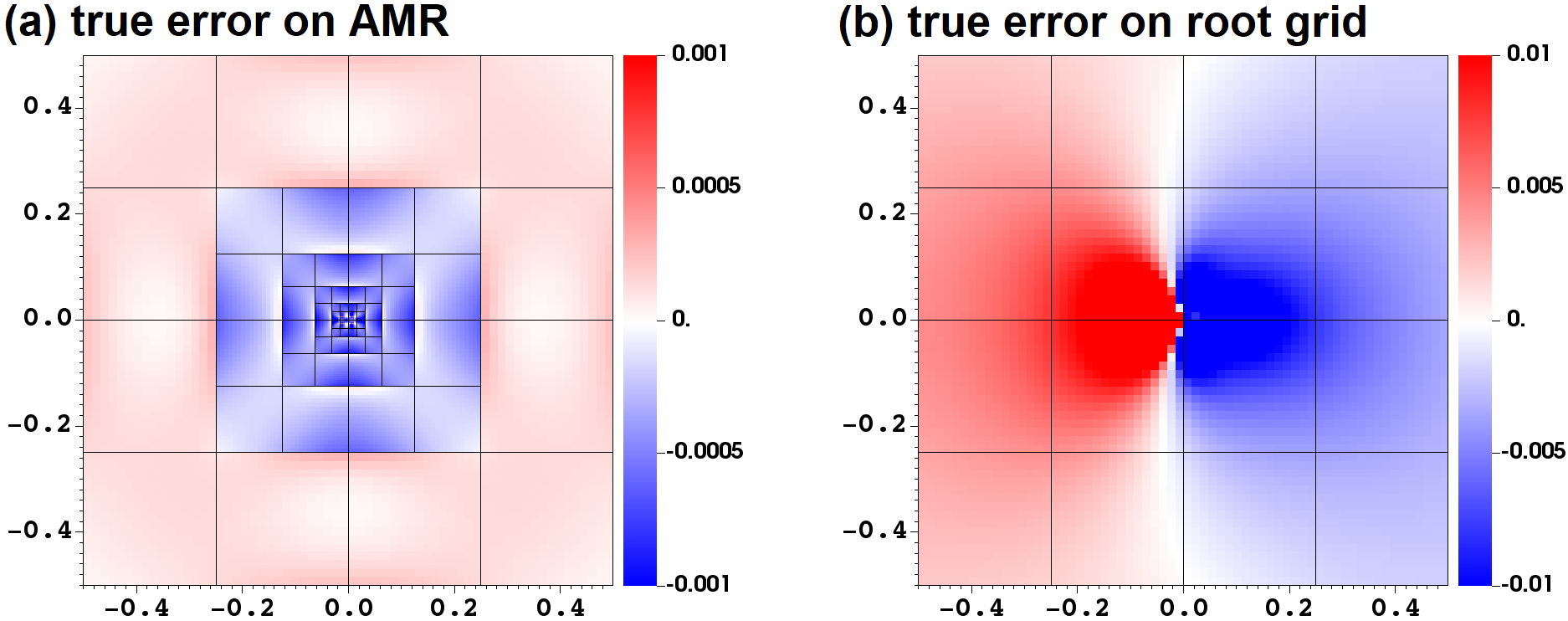}
\end{center}
\caption{Distribution of the normalized error against the analytic solution of the binary test problem (equation \ref{eq:normtrueerr}). (a) The fully-converged solution on the AMR grid. (b) The fully-converged solution obtained using only the root level. Black lines show boundaries between \texttt{MeshBlocks}. The range of the color bar in Panel (b) is ten times larger than that in Panel (a).}
\label{fig:amr_lbl}
\end{figure*}
In this Appendix, we present a rough comparison between our self-gravity solver and the level-by-level approach used in other simulation codes \citep[e.g.][]{art,ramsesmg,enzo,dispatchgrav}. In the latter method, solution on each level is calculated separately using solutions on coarser levels as boundary conditions for finer levels. This method is also referred to as one-way coupling. In contrast, our solver calculates the potential consistently across all AMR levels.

In order to demonstrate the difference between these approaches qualitatively, we present a very simple experiment instead of implementing the level-by-level approach for ourselves and comparing results side-by-side. Using the binary test problem presented in Section~\ref{sec:binary} with the root level resolution of $h=1/64$, we compare fully-converged solutions obtained using AMR and using the root level only. The latter is corresponding to the coarsest level solution used in the level-by-level approach. Figure~\ref{fig:amr_lbl} shows distributions of normalized true error (equation \ref{eq:normtrueerr}). Here we compare the error on the root grid (the region outside the refined \texttt{MeshBlocks}). With AMR, the error in this region is only about a few $\times 10^{-4}$ (Panel (a)). On the other hand, as shown in Panel (b), the error of the solution calculated with the root grid only is as large as 1\%.

In the level-by-level approach, this solution on the root level is interpolated to the finer level and used as the boundary condition. Therefore, the solution on the finer level should also contain error of the same order, and the error propagates further to the finer levels by repeating this procedure. As a result, it is expected that the solution obtained with the level-by-level approach is less accurate by more than one order of manitude and contains a percent-level error. Of course, this is a rough estimate in this particular setup and actual error depends on problems, including the grid configuration and the number of AMR levels. In this particular problem, the binary is not well resolved on the root level in this setup. The error can be smaller if the density distribution is well resolved even on the coarser levels, but in principle the level-by-level approach should suffer from this kind of error. \citet{wangyen2020} proposed an improved scheme for the interpolation at level boundaries, but it does not fundamentally resolve this issue. It is worth noting that this error in the level-by-level approach is not random but systematic, and it may cause non-negligible error in long-term simulations. For example, if there is a binary system not well resolved on the coarser levels, there can be substantial error in the torque produced by the binary potential, and may affect the angular momentum evolution of the system. With the level-by-level approach, AMR grid should be (more) carefully constructed in order to control the error. This result indicates that our implementation is more accurate and consistent compared to the level-by-level approach. 

\end{document}